\documentclass[11pt]{article}
\pdfoutput=1
\usepackage{jheppub} 
\usepackage{amsmath,amssymb,epsfig,amsfonts}
\usepackage{epsf}
\usepackage{makeidx}
 \usepackage{slashed}
    \usepackage{verbatim} 
    \usepackage{float}
    \restylefloat{table}
    \usepackage{pdflscape}
    \usepackage{tikz}
    \usepackage{pifont}
\usepackage{array}
    \usepackage{youngtab}
    \usepackage{multirow}
    \usepackage{xcolor}
    \usepackage{ulem}
    \usepackage{cleveref}
    \usepackage{tensor}
    \usepackage{todonotes}
    \usepackage{mathrsfs}
    \usepackage{changepage}
    \usepackage{anyfontsize}
   \usepackage{caption}
\usepackage{orcidlink}

\makeatletter

\newcommand{\be}{\begin{equation}}
\newcommand{\ee}{\end{equation}}
\newcommand{\ba}{\begin{aligned}}
\newcommand{\ea}{\end{aligned}}

\newcommand{\dd}{\mathrm{d}}

\newcommand{\me}{\mathrm{e}}
\newcommand{\ii}{\mathrm{i}}
\newcommand{\vol}{\mathrm{vol}}
\newcommand{\Vol}{\mathrm{Vol}}

\newcommand{\ls}{\ell_s}

\newcommand{\del}{\partial}
\newcommand{\nn}{\nonumber}
\newcommand{\la}{\langle}
\newcommand{\ra}{\rangle}
\newcommand{\hook}{\mathbin{\rule[.2ex]{.4em}{.03em}\rule[.2ex]{.03em}{.9ex}}}

\makeatother

\begin{document}

\baselineskip=18pt  
\numberwithin{equation}{section}  
\allowdisplaybreaks  


%
%


\thispagestyle{empty}


\vspace*{1cm} 
\begin{center}
{\fontsize{16pt}{23pt}\selectfont\textbf{A geometric dual of F-maximization in massive type IIA}\vspace{10mm}}
 
 \renewcommand{\thefootnote}{}
\begin{center}
 \fontsize{14pt}{23pt}\selectfont{Christopher Couzens\textsuperscript{\orcidlink{0000-0001-9659-8550}}\footnotetext{\href{mailito:christopher.couzens@maths.ox.ac.uk}{christopher.couzens@maths.ox.ac.uk}} and  Alice L\"uscher\textsuperscript{\orcidlink{0009-0001-8231-3080}}\footnotetext{\href{mailito:alice.luscher@maths.ox.ac.uk}{alice.luscher@maths.ox.ac.uk}}}

\end{center}
\vskip .2cm

 \vspace*{.5cm} 
 \fontsize{12pt}{14pt}\selectfont{{ Mathematical Institute, University of Oxford,\\ Andrew Wiles Building, Radcliffe Observatory Quarter,\\ Woodstock Road, Oxford, OX2 6GG, U.K.}}

 {\tt {}}

\vspace*{0.8cm}
\end{center}

 \renewcommand{\thefootnote}{\arabic{footnote}}
 
\begin{center} {\bf Abstract } 
\end{center}
 Using equivariant localization we construct a geometric dual of F-maximization in massive type IIA supergravity. Our results use only topological data to quantize the fluxes, compute the free-energy and conformal dimensions of operators in the dual field theory without the need for explicit solutions. We utilize our formalism to study various classes of solutions, including examples where an explicit solution is not known.

\vspace*{5.5cm}

\begin{flushright}
{\emph{Dedicated to Alan Ernest Couzens}}\\
\end{flushright}

\newpage

\tableofcontents
\printindex

\newpage
\section{Introduction}


Finding metrics satisfying the non-linear Einstein equations is a notoriously difficult problem. Naively, this seems to pose a computational problem to computing observables holographically. However, recent advances in defining extremal problems in gravity allows one to extract out certain holographic observables. These geometric extremal problems have a corresponding field theoretic dual, whether it be: a-maximization \cite{Intriligator:2003jj}; F-extremization, \cite{Jafferis:2010un}; c-extremization \cite{Benini:2012cz,Benini:2013cda}, or $\mathcal{I}$-extremization \cite{Benini:2015eyy}.  This program of studying geometric extremal problems began with the geometric dual of a-maximization/F-extremization in \cite{Martelli:2006yb,Martelli:2005tp} and more recently has been extended to $\mathcal{I}$/c-extremization \cite{Couzens:2018wnk} see also \cite{Hosseini:2019use,Hosseini:2019ddy,Kim:2019umc,Gauntlett:2019pqg,Gauntlett:2019roi,Gauntlett:2018dpc}. 

Recently in \cite{BenettiGenolini:2023kxp,BenettiGenolini:2023ndb} equivariant localization was used to setup such extremal problems in a variety of different setups \cite{BenettiGenolini:2023yfe,BenettiGenolini:2024kyy,Suh:2024asy}.\footnote{See also \cite{Martelli:2023oqk,Colombo:2023fhu} for a complementary but distinct approach and \cite{BenettiGenolini:2019jdz} for an earlier incarnation of localization. } This allows one to compute physical observables of the dual SCFT without knowing the full supergravity solution. This paper applies these methods to (massive) type IIA on AdS$_4\times M_6$. Such solutions have been classified in \cite{Passias:2018zlm}, see also \cite{Lust:2009mb}. Concretely we use equivariant localization to perform flux quantization, compute the holographic free energy, and conformal dimensions of certain BPS operators. We use the formalism to derive certain gravitational block formulae conjectured in the literature \cite{Faedo:2022rqx}, reproduce known results and make predictions for new solutions. 

At first sight, given that our internal manifold $M_6$ shares many similarities with the internal manifold of AdS$_5$ solutions in M-theory \cite{Gauntlett:2004zh} which were equivariantly localized in  \cite{BenettiGenolini:2023ndb}, this may seem a trivial extension of their work. We find that although there are similarities there is also a rich structure of possible internal geometries which are not possible in the AdS$_5$ M-theory solutions. In particular one novel aspect of our work is the inclusion of boundaries on the internal space on which we localize. These boundaries arise in our construction from the presence of brane sources which cap off the space. One finds that such singular geometries can also be localized and this opens up a wide avenue of solutions where these techniques can be applied. 

AdS$_4$ solutions of (massive) type IIA supergravity preserving $\mathcal{N}=2$ supersymmetry with an SU$(2)$ structure were classified in \cite{Passias:2018zlm}. The R-symmetry of the putative dual field theories is (at least) U$(1)$, ergo, via the usual AdS/CFT lore there is a U$(1)$-isometry of the metric and an associated Killing vector field. We will use this U$(1)$ isometry to localize integrals in our setup. Given the geometric setup in \cite{Passias:2018zlm}, using brute force, we construct a set of equivariantly closed polyforms. To contruct these polyforms we use a subset of the torsion conditions. Since only a subset are imposed we are to some extent "off-shell" and one needs to extremize over the free parameters to find the on-shell results. This provides a geometric interpretation to F-maximization principles of the dual SCFTs \cite{Jafferis:2010un,Jafferis:2011zi,Closset:2012vg}. A holographic approach to F-maximization was already discussed in \cite{Freedman:2013oja,Fluder:2015eoa} for Sasaki--Einstein geometries and a class of massive type IIA solutions. In this work we extend this geometric extremal problem to all $\mathcal{N}=2$ AdS$_4$ solutions in (massive) type IIA with an SU$(2)$ structure. 

The plan of the paper is as follows. In section \ref{sec:localizationreview} we briefly review equivariant localization and the Atiyah--Bott--Berline--Vergne localization theorem. In particular we discuss the contributions from boundaries. In section \ref{sec:IIAall} we review the solutions of \cite{Passias:2018zlm}, study the various options for constructing a well-defined background using O8 sources and then construct various equivariantly closed polyforms and present the general localized integrals for these polyforms. In section \ref{sec:examples} we use our results to study five distinct examples of possible geometries, each with different behaviours. We conclude in section \ref{sec:conclusion}. Some technical material on cohomology relations is relegated to appendix \ref{app:cohomology}.

\section{Review of equivariant localization}\label{sec:localizationreview}

Before we start to apply localization to our setup it is instructive to first review the formalism. One additional point that we wish to emphasize is the modification of the usual localization formulae in the presence of a boundary. This section closely follows \cite{BenettiGenolini:2023kxp, BenettiGenolini:2023ndb}, which initiated the idea of applying localization to supergravity, adding in a discussion on boundary contributions. See also \cite{berline2003heat,Cremonesi:2013twh} for an introduction to equivariant localization. The reader familiar with these techniques can safely skip to section 3. 

Consider a $d$-dimensional space $M$ with a U$(1)$ Killing vector $\xi$.\footnote{We will specialize to $d=6$ in the following though the discussion below holds more generally. Further one could replace the U$(1)$ action with any compact Lie group action with minimal modifications.} The equivariant exterior derivative is defined to be
\begin{equation}
\dd_\xi = \dd - \xi\hook\,,
\end{equation}
and acts on polyforms. It has the property that $\dd_\xi^2=-\mathcal{L}_\xi$ (the Lie derivative), such that it defines an equivariant cohomology on the space of invariant polyforms $\Phi$.
A polyform is said to be equivariantly closed if $\dd_\xi\Phi=0$.
The integral over an invariant even dimensional submanifold $\Gamma\subset M$ of an equivariantly closed polyform $\Phi$ can be evaluated using the BV--AB theorem \cite{BV:1982,Atiyah:1984px}, which states that such an integral localizes to the fixed point set of the group action.  
Explicitly let us denote by $\Sigma\subset \Gamma$ a fixed submanifold of $\xi$ (i.e.\ $\xi=0$ on $\Sigma$) of codimension $2k$ and $f:\Sigma\hookrightarrow \Gamma$ the embedding of the fixed point locus, then the BV--AB theorem gives
\begin{equation}\label{eq:BVABgen}
\int_{\Gamma}\Phi=\sum_{\Sigma}\frac{1}{d_{2k}}\int_{\Sigma} \frac{f^{*}\Phi}{e_{\xi}(\mathcal{N})}\, ,
\end{equation}
where $e_{\xi}(\mathcal{N})$ is the Euler form of the normal bundle and $d_{2k}$ is the order of the orbifold structure group of $\Sigma$, meaning that for $N_\Sigma=\mathbb{R}^{2k}/G$ locally, $d_{2k}\in\mathbb{N}$ is the order of the finite group $G$.

We may simplify the BV--AB formula under the assumption that the normal bundle $N_\Sigma$ of $\Sigma$ in $\Gamma$ decomposes as a sum of line bundles $N_\Sigma=\oplus_{i=1}^k L_i$.\footnote{This assumption is satisfied by almost all of the examples considered in this paper. It is used to write the Euler class of $N_\Sigma$ in terms of first Chern class in the formula \eqref{BVAB}. This more general formula will be important in section \ref{sec:SE} where such a decomposition in terms of line bundles is not possible.}  Then the BV--AB formula, with this splitting assumption reduces to
\be\label{BVAB}
    \int_\Gamma \Phi= \sum_\Sigma \frac{1}{d_{2k}}\frac{(2\pi)^k}{\prod_{i=1}^k\epsilon_i}\int_\Sigma \frac{f^*\Phi}{[1+\frac{2\pi}{\epsilon_i}c_1(L_i)]}\,,
\ee
where $\epsilon_i$ are the weights of $\xi$ on $N_\Sigma$, i.e.\ in local coordinates 
$    \xi = \sum_{i=1}^k\epsilon_i\del_{\varphi_i}
$
with $\del_{\varphi_i}$ rotating $L_i$, and $c_1(L_i)$ are the first Chern classes. 
In practice the denominator is to be expanded in a power series in the Chern class, as illustrated below, which truncates at order dim$(\Sigma)$ at most. 
In summary, the game to be played is to build equivariantly closed polyforms whose top form is a physical quantity that we want to measure. The physical observable, obtained by integrating the top form over some (sub-)manifold of our internal space, is then evaluated using the BV--AB formula. This gives the observable as a sum of contributions of the lower forms evaluated on fixed submanifolds.

To make  the BV--AB formula a little more explicit let us see how the general fixed point formula \eqref{BVAB} reads for the cases of interest in this paper. The internal space is a 6 dimensional compact manifold $M_6$, and we will compute integrals over 2 and 4 cycles as well as the full space. The BV--AB formula then gives:
\begin{align}\label{BVAB246}
    \int_{\Gamma_{2}} \Phi_2 =  &\sum_{\Sigma_0} \frac{2\pi}{d_0}\frac{\Phi_0}{\epsilon_1}\Big|_{\Sigma_0} \,,\\ \nn
    \int_{\Gamma_4} \Phi_4 = &\sum_{\Sigma_0} \frac{(2\pi)^2}{d_0}\frac{\Phi_0}{\epsilon_1\epsilon_2}\Big|_{\Sigma_0}+ \sum_{\Sigma_2} \frac{2\pi}{d_2}\int_{\Sigma_2}\left[\frac{\Phi_2}{\epsilon_1}-\frac{2\pi\Phi_0}{\epsilon_1^2}c_1(L)\right]\,, \\ \nn
    \int_{M_6} \Phi_6 = &\sum_{\Sigma_0} \frac{(2\pi)^3}{d_0}\frac{\Phi_0}{\epsilon_1\epsilon_2\epsilon_3}\Big|_{\Sigma_0}+ \sum_{\Sigma_2} \frac{(2\pi)^2}{d_2}\frac{1}{\epsilon_1\epsilon_2}\int_{\Sigma_2}\left[\Phi_2-2\pi\Phi_0\Big(\frac{c_1(L_1)}{\epsilon_1}+\frac{c_1(L_2)}{\epsilon_2}\Big)\right] \\ \nn  +  &\sum_{\Sigma_4}  \frac{2\pi}{d_4}\int_{\Sigma_4}\left[\frac{\Phi_4}{\epsilon_1}-\frac{2\pi\Phi_2}{\epsilon_1^2}\wedge c_1(L)+\frac{(2\pi)^2\Phi_0}{\epsilon_1^3}c_1(L)\wedge c_1(L)\right]\,.
\end{align}

\subsection{Boundary contributions}

As we will see shortly, in our setup the internal manifold can have boundaries and this requires an extension of the BV--AB formula \eqref{BVAB}. We make the restriction that the boundary of the manifold does not have any fixed points and let $f:\partial M \hookrightarrow M$ be the embedding of the boundary into $M$. Then to the BV--AB formula in \eqref{BVAB} we add in an additional boundary contribution \cite{Szabo:1996md}:
\begin{equation}\label{eq:BVABboundary}
  \int_{\Gamma}\Phi= \text{BV--AB\eqref{BVAB}} -\int_{\partial \Gamma}f^*\frac{\xi \wedge \Phi}{\dd_{\xi}\xi}\, .
\end{equation}
We understand the inverse of the polyform $\dd_{\xi} \xi$ via a formal geometric series, 
\begin{equation}
    \frac{1}{\dd_{\xi}\xi}= -\frac{1}{|\xi|^2(1-|\xi|^{-2} \dd\xi)}=-\frac{1}{|\xi|^2}\bigg[1+\sum_{a=1}^{\infty}\left(\frac{\dd \xi}{|\xi|^2}\right)^{a}\bigg]\, .
\end{equation}
It is obvious from the formula that the boundary must not have fixed points of the U$(1)$ given the factors of $|\xi|^2$ appearing in the denominator. This was a choice that we made and there exists a more general formula when this is not true, \cite{Szabo:1996md}.

To better explain the contributions from the boundary let us study an example. Consider a four-sphere and let us put in an arbitrary boundary which preserves a choice of U$(1)$ action that we will specify shortly. Let us take the following metric on the four-sphere\footnote{We use the same conventions as in \cite{BenettiGenolini:2023ndb} in order to use their polyforms.}
\begin{equation}
    \dd s^2= \dd \alpha^2+\sin^2\alpha\big(\dd\theta^2+\sin^2\theta \dd\phi_1^2+\cos^2\theta \dd\phi_2^2\big)\, .
\end{equation}
For the round four-sphere the ranges of the coordinates are: $\alpha\in [0,\pi]$, $\theta\in[0,\tfrac{\pi}{2}]$, and $\phi_1,\phi_2\in [0,2\pi]$. We will localize using the Killing vector
\begin{equation}
    \xi=b_1\partial_{\phi_1}+b_2\partial_{\phi_2}\, ,
\end{equation}
with both $b$'s non-zero. The norm of this Killing vector field is 
\begin{equation}
    |\xi|^2=\sin^2\alpha( b_1^2 \sin^2\theta+b_2^2\cos^2\theta)\, ,
\end{equation}
which clearly vanishes at the poles of the $S^4$ at $\alpha=0,\pi$. Rather than taking the round four-sphere we will restrict $\alpha$ to $\alpha\in [0,\alpha_0]$ with $0<\alpha_0<\pi$ and call the space $X_{\alpha_0}$. For $\alpha_0=\tfrac{\pi}{2}$ this is the usual four-dimensional hemi-sphere. It is simple to explicitly compute the volume for this metric and we find
\begin{equation}\label{eq:volXalpha0}
    \Vol(X_{\alpha_0})=\frac{4\pi^2}{3}+ \frac{2\pi^2\cos\alpha_0(\cos\alpha_0^2-3)}{3}\, .
\end{equation}

We now want to reproduce this using our localization formulae. We have a fixed point at the pole and a boundary term at $\alpha=\alpha_0$. Using the results in \cite{BenettiGenolini:2023ndb} the equivariantly closed polyform for the (weighted) volume form is given by  $\Phi=\Phi_4+\Phi_2+\Phi_0$ with:
\begin{equation}
    \begin{split}
        \Phi_4&=\frac{1}{b_1b_2}\vol=\frac{1}{b_1 b_2}\sin^3 \alpha \sin\theta\cos\theta \dd\alpha\wedge \dd\theta\wedge \dd\phi_1\wedge \dd\phi_2\, ,\\
        \Phi_2&= -\frac{1}{2}\sin^3\alpha \left(\frac{1}{b_1} \sin^2\theta \dd\phi_1+\frac{1}{b_2}\cos^2\theta \dd\phi_2\right)\wedge \dd\alpha\, ,\\
        \Phi_0&=\frac{1}{6}(3\cos\alpha-\cos^3\alpha)\, .
    \end{split}
\end{equation}
The weighted volume is then
\begin{equation}
\begin{split}
    \int_{X_{\alpha_0}}\frac{1}{b_1 b_2}\vol(X_{\alpha_0})&=\frac{(2\pi)^2}{b_1 b_2}\Phi_0\Big|_{\alpha=0}+\int_{\partial X_{\alpha_0}} f^{*}\frac{1}{|\xi|^2}\left[ \xi\wedge \Phi_2+\frac{\Phi_0}{|\xi|^2} \xi\wedge \dd \xi\right]\\
    &=\frac{4\pi^2}{3 b_1 b_2}+\int_{\partial X_{\alpha_0}}\frac{b_1 b_2 \cos\alpha_0(3-\cos^2\alpha_0)\cos\theta \sin\theta }{3(b_1^2 \sin^2\theta +b_2^2 \cos^2\theta)^2}\dd\theta \wedge \dd\phi_1\wedge \dd\phi_2
\end{split}
\end{equation}
Note that the contribution only arises from the second term involving $\Phi_0$ since $f^*\Phi_2=0$. We now need to perform the integral over $\partial X_{\alpha_0}$. Note that this is just an integral over a round three sphere and one could in principle use equivariant localization again to perform this integral. We will refrain from doing this but it is not difficult to see that the final result is:
\begin{equation}
  \int_{X_{\alpha_0}}\frac{1}{b_1 b_2}\vol(X_{\alpha_0})=\frac{2\pi^2(2-3 \cos\alpha_0+\cos^3\alpha_0)}{3b_1 b_2}\, ,  
\end{equation}
and therefore we find that we recover the correct result in \eqref{eq:volXalpha0}. It is important to note that for generic $\alpha_0\neq\tfrac{\pi}{2}$ there is a contribution from the boundary. One expects that it is possible to use equivariant localization to also evaluate the integrals for the boundary contributions and this would be an interesting problem to study in the future. 

Having given the generic formula with boundaries it turns out that the boundary contributions from the observables we will consider vanish because $f^*\Phi=0$ for our integrals. This is not a generic feature, indeed the example above had a boundary contribution, except if we looked at the hemi-sphere. Rather, it is special to the various setups that we consider and the types of boundary there. One expects that by looking at different admissible boundary conditions or more refined observables, boundary contributions will no longer vanish. We hope to study this in the future.

\section{General \texorpdfstring{AdS$_4$}{AdS(4)} solutions of massive type IIA}\label{sec:IIAall}

$\mathcal{N}=2$ preserving AdS$_4$ solutions of massive type IIA with an SU$(2)$ structure were classified in \cite{Passias:2018zlm}. It was shown that there are two distinct classes, named class K (K\"ahler) and class HK (hyper-K\"ahler). Both classes are topologically an $S^2$ bundle over a four-dimensional space $M_4$ which is K\"ahler or hyper-K\"ahler respectively. In both cases the 10d metric takes the form
\be
\dd s^2_{10}=\me^{2A}\left[\dd s^2_{\text{AdS}_4}+\dd s^2_{M_6}\right]\,,\label{eq:10dmet}
\ee
with $M_6$ dependent on the class. The class K solutions will be the main focus of this work and we will ignore the class HK solutions since the solutions in the HK class are essentially unique and therefore localization is not needed to compute observables in these theories. We work in conventions where the length scale on AdS$_4$ is set to 1. This may be reinstated with a little dimensional analysis.

\subsection{Setup}

The internal metric for solutions of class K takes the form
\be\label{eq:6dmet}
\dd s^2_{M_6}=\frac{1}{\me^{4A}-y^2}\dd y^2+\frac{1}{4}(1-\me^{-4A}y^2)D \psi^2 +\frac{y}{F_0^2\me^{4A}y^2 +l^2} g^{(4)}_{ij}(y,x)\dd x^i\dd x^j\,,
\ee
with $g^{(4)}$ a four-dimensional K\"ahler metric at fixed $y$ coordinate, $D\psi=\dd\psi+\rho$. The SU$(2)$ structure forms on $M_4$ satisfy:\footnote{In the original paper \cite{Passias:2018zlm} the $j$ used here is denoted by $\hat \jmath$ and is normalized differently from $j$ there. }
\begin{align}
\partial_{\psi}j&=0\, ,\quad &\partial_y j&=\frac{1}{2}(F_0^2 y^2 +l^{2} y^{-2})\dd_4 \rho\, ,&\dd_4 j&=0\, ,\\
\partial_{\psi}\omega&=0\, ,\quad &\partial_y\omega&=-\frac{1}{2}(F_0^2 y^2 +l^2 y^{-2}) T \omega\, ,\quad &\dd_4 \omega&=\ii P\wedge \omega\, ,
\end{align}
where
\begin{align}
P&\equiv-\rho +\ii \frac{2\me^{4A}( F_0^2 y^4+l^2)}{(\me^{4A}-y^2)(F_0^2\me^{4A}y^2+l^2 )}\dd_4 A\, ,\\ \label{T}
T&\equiv \frac{\partial_y(\me^{4A}y^2)}{(\me^{4A}-y^2)(F_0^2\me^{4A}y^2+l^2)}\, .
\end{align}
There are two constant parameters, $F_0$ the Romans mass and $l$. It is necessary that at least one of these parameters is non-zero. A non-zero value of $F_0$ signifies the presence of D8-branes in the setup. On the other hand the presence of a non-zero $l$ signifies the presence of D2- and D6-branes as can be seen more clearly by considering the fluxes supporting the solution. Note that if both are non-zero by a redefinition of the $y$ coordinate and rescaling of the metric and fluxes $l$ may be set without loss of generality to a non-zero value, such as $l=1$.

The flux quantization of the Romans mass $F_0$ imposes
\begin{equation}
    n_0=2\pi\ell_s F_0\in \mathbb{Z}\,.
\end{equation}
The magnetic parts of the RR-fluxes are:
\begin{equation}\label{RRfluxes}
    \begin{split}
        F_2&=l\left(\frac{1}{2y}\dd \rho-\frac{1}{2}\dd(\me^{-4 A}yD\psi)+\frac{1}{F_0^2\me^{4A} y^2 +l^2}j\right)\, ,\\
        F_4&=-\frac{F_0}{2}\Bigg[\frac{\me^{4A}y^2}{(F_0^2 \me^{4A}y^2+l^2)^2}j\wedge j \\
        &-\Big(\dd(\me^{-4A}y D\psi)- y^{-1}\dd\rho\Big)\wedge \left(\frac{\me^{4A}y^2}{F_0^2 \me^{4A}y^2+l^2}j +\frac{y^2}{2}\dd y\wedge D\psi\right)\Bigg]\, ,\\
        F_6&= \frac{3l}{4} \frac{ y^2} {(F_0^2\me^{4A} y^2 +l^2)^2}\dd y \wedge D\psi \wedge j\wedge j\, .
    \end{split}
\end{equation}
For the NS-NS sector  it is useful to give two different expressions for the NS-NS two-form related by gauge transformations:
\begin{equation}\label{eq:Bfield}
    \begin{split}
        B_1&=-\frac{F_0}{l}\left( \frac{\me^{4A} y^2}{F_0^2\me^{4 A}y^2+l^2}j+\frac{y^2}{2}\dd y \wedge D\psi\right)\, ,\\
        B_2&=\frac{l}{F_0}\left(\frac{1}{F_0^2\me^{4 A}y^2+l^2}j+\frac{1}{2 y^2}\dd y \wedge D\psi\right)\, .
    \end{split}
\end{equation}
Note that the former is ill-defined in the $l\rightarrow 0$ limit while the latter is ill-defined in the $F_0\rightarrow 0$ limit so one should choose the correct gauge when specialising to either case. 
Finally the dilaton is given by:
\begin{equation}\label{eq:dil}
    \me^{2\phi}=\frac{\me^{6A}}{F_0^2 \me^{4A}y^2+l^2 }\, ,
\end{equation}
which makes manifest the necessity to have at least one of the parameters $F_0$ or $l$ non-zero.

The above fluxes are the (magnetic) ones which appear in the equations of motion. They, however, are not the ones one would use to quantize the fluxes. These are the Page fluxes: $f=F\wedge\me^{-B}$ in polyform notation. Here $F$ is the polyform of magnetic fluxes, $B$ is either of the representatives of $B$ given in \eqref{eq:Bfield} and one extracts out the part of form degree $p$. Recall that Page fluxes are closed, hence define conserved charges, but are not gauge invariant. The quantization condition reads
\begin{equation}
    \frac{1}{(2\pi \ls)^{p-1}}\int_{\Sigma_p}f_p\in \mathbb{Z}\, .
\end{equation} 
In the general class ($l\neq0$) the Page fluxes read:
\begin{equation}
\begin{split}
f_2&= \frac{l}{2}\dd\Big[y^{-1}\big(1-\me^{-4 A}y^2\big)D\psi\Big]- F_0 (B_1-B_2)\, ,\\
f_4&=\frac{F_0}{2 l^2}\Big[\frac{\me^{4A} y^2}{F_0^2 \me^{4A} y^2+l^2}j\wedge j + y^2 \dd y\wedge D\psi \wedge j\Big]\, ,\\
f_6 &=\frac{y^2}{4 l^3}\frac{F_0^2\me^{4A} y^2+3 l^2}{F_0^2\me^{4A} y^2+l^2 }\dd y \wedge D\psi \wedge j \wedge j\, .
\end{split}
\end{equation}
Observe that these fluxes do not admit a well-defined $l=0$ limit, though one can safely take the $F_0=0$ limit. If $l=0$ one notices that both $F_2$ and $F_6$ vanish identically and that one may pick a gauge in which $B=0$ too, i.e.\ pick $B=B_2$. It follows that the only non-trivial flux is $F_4$ and its charge defines a conserved quantity. We must therefore specify as initial data whether $l=0$ or not and we will explicitly state this when relevant. In section \ref{sec:genericpoly} we will construct equivariantly closed polyforms for $l\neq0$ and in section \ref{sec:l=0} the analogous polyforms for the $l=0$ case.

\subsection{Degenerations and global regularity}\label{sec:degeneration}

Before we set up the localization problem we need to first discuss the possible global completions of the metric. It turns out that there is a very rich structure for the class K solutions. In order to have well defined, globally complete (and compact)\footnote{One can easily release the compact condition by studying solutions corresponding to 3d conformal defects in a higher dimensional parent theory, see for example \cite{Gutperle:2022pgw,Gutperle:2023yrd,Capuozzo:2023fll} for solutions of this form in M-theory. We will not pursue this here and thank Pieter Bomans for discussions on this topic.} solutions we need to fix the length of the line interval with coordinate $y$. The usual way, amenable to localization, is for the localizing circle to shrink at some value of $y$. Depending on the shrinking of the remaining part of the internal metric one obtains either fixed points or fixed surfaces of co-dimension 2 or 4.

There is however another way in which to globally complete the metric: the presence of brane singularities. Chosen correctly these cap off the space, giving a global completion, despite the localizing circle remaining of non-zero size. From the localization viewpoint these lead to boundaries in the internal space that we must localize over. In this section we will study one singular brane solution that we can have which cap off the space in a well-defined manner with a physical interpretation. Many of the results can be obtained from \cite{Passias:2018zlm}, however we rephrase the conditions in terms of the behaviour of the circle at the special values of $y$. There are many other options that one can consider involving (possibly smeared) orientifolds, see for example \cite{Passias:2018zlm}, however we will content ourselves with this single choice as it is quite universal.

We will be interested in capping off the spacetime with an O8-plane singularity with divergent dilaton. The metric and dilaton behaviour of such an O8 singularity is
\begin{equation}\label{eq:O8plane}
    \begin{split}
        \dd s^2&\sim \frac{1}{\sqrt{r}}\dd s^2_{9}+\sqrt{r}\dd r^2\, ,\\
        \me^{2\phi}&\sim r^{-5/2}\, ,
    \end{split}
\end{equation}
where we have kept only the leading order pieces and the plane is located at $r=0$. Note that there are other choices of O8-plane which would not have the above divergence structure, this choice is taken to match with the O8-plane in the Brandhuber--Oz solution \cite{Brandhuber:1999np} that will be relevant later.

We can now study the existence of such a degeneration for the metric in \eqref{eq:10dmet} and dilaton in \eqref{eq:dil}.  
It is not difficult to show that this is only possible for $l=0$ and $y=0$. The argument is simple, one immediately sees that the warp factor must diverge as $\me^{2A}\sim r^{-1/2}$. Requiring that the dilaton has the correct divergence fixes $l=0$. Finally requiring that the metric in \eqref{eq:6dmet} degenerates correctly fixes $y=0$ to be the location of the branes, that is it identifies $y$ with $r$ in \eqref{eq:O8plane}.   

In conclusion to have an O8-plane it follows that $l=0$ and as $y\rightarrow 0$ the warp factor must diverge as
\begin{equation}
    \me^{2A}\sim y^{-1/2}\, .
\end{equation}
Around $y=0$ the metric and dilaton have the following behaviour
\begin{equation}
    \begin{split}
        \dd s^2&\sim \frac{1}{\sqrt{y}}\bigg(\dd s^2_{\text{AdS}_4}+\frac{1}{4}D\psi^2+\frac{1}{F_0^2}g_{ij}^{(4)}(0,x)\dd x^i \dd x^j\bigg)+\sqrt{y}\dd y^2\, ,\\
        \me^{2\phi}&\sim y^{-5/2}\, .
    \end{split}
\end{equation} 
One needs to further refine the warp factor divergence to the next order. In general expanding the warp factor around $y=0$ we have
\begin{equation}
    \me^{2 A}\sim \frac{a_1}{\sqrt{y}}+a_2 \sqrt{y}\, .
\end{equation}
For $a_2=0$ it follows that $F_4$ as given in \eqref{eq:F4l0} vanishes at $y=0$. For $a_2\neq0$ we find that there is a term proportional to the volume form on $M_4$. Comparing to the explicit solutions in \cite{Passias:2018zlm} $a_2$ plays the role of the parameter $\sigma$ in section 4.1.4 there. We see that the value of $a_2$ changes the homology relation we will impose later since the flux does not vanish on the boundary of $M_6$.
We reiterate that this is not the only option of a brane singularity that we may use to cap off the space however this is the most interesting choice  in order to connect to the compactification of 5d SCFTs on a two-dimensional surface.

\subsection{Polyforms}

In order to perform equivariant localization we need to first pick an action on which we wish to localize. There is an obvious U$(1)$-action given by the R-symmetry Killing vector $\xi=\partial_{\psi}$.\footnote{Observe that we have taken a different normalization for $\xi$ to that in \cite{Passias:2018zlm} for which $\xi_{\text{there}}=4\xi_{\text{here}}$.} The Killing vector $\xi=\del_\psi$ satisfies $\xi\hook D\psi=1$, and contractions into the SU$(2)$ structure forms vanish: $\xi\hook j=\xi\hook \omega=0$. As reviewed in section \ref{sec:localizationreview}, we want to construct various equivariantly closed polyforms i.e.\ polyforms $\Phi$ which satisfy $\dd_\xi\Phi\equiv(\dd-\xi\hook)\Phi=0$. This allows us to localize integrals of the top-form using the BV--AB fixed point formula. In this section we will construct these polyforms and give generic formulae for the BV--AB theorem applied to these polyforms in section \ref{sec:loc} before using them later in section \ref{sec:examples} to study various choices for $M_6$. In constructing it is necessary to separate the two cases of $l\neq 0$ and $l=0$, though one can safely set $F_0=0$ in the former case. 

There are a number of different equivariantly closed polyforms that we want to construct. Firstly, we need to perform flux quantization and therefore we need to equivariantly complete the Page fluxes, secondly, we want to compute the free-energy of the various solutions, this is computed via:
\begin{equation}
\mathcal{F}=\frac{16\pi^3}{(2\pi \ls)^8}\int_{M_6} \me^{8A-2\phi}\vol_{M_6}\,.
\end{equation}
We therefore need to find a polyform with top-component $\me^{8A-2\phi}\vol_{M_6}$. 
Finally, there are various probe branes which we can consider. These wrap calibrated cycles in the internal space and depending on the type of cycle wrapped give rise to observables in the three-dimensional dual SCFT such as conformal dimensions of BPS particles. We will focus on D2-branes wrapping calibrated two-cycles \cite{Bah:2018lyv} though it would be interesting to consider other probe branes in the geometry.

\subsubsection{The \texorpdfstring{$l\neq 0$}{l ≠ 0} case}\label{sec:genericpoly}

After a slightly tedious, but straightforward computation, one can compute all the various polyforms outlined above. We suppress the details of the computations, presenting just the final polyforms. 
The polyform for the free energy is 
\begin{equation}
\begin{split}
\Phi^{\mathcal{F}}&= \me^{8A-2\phi}\star_1-\frac{1}{12}\left(\frac{y^3}{F_0^2 \me^{4A}y^2+l^2} j\wedge j + y \dd y \wedge D\psi \wedge j\right)\\
&+\frac{1}{24}\left( y^2 j +\frac{1}{2}(F_0^2 y^4+l^2) \dd y\wedge D\psi\right)
-\frac{y}{48}\left(\frac{F_0^2 y^4}{5}+l^2  \right)\, .
\end{split}
\end{equation}
For the Page fluxes we find:
\begin{equation}
    \begin{split}
        \Phi^{f_2}&= f_2 -\frac{y}{6 l}(F_0^2 y^2-3 l^2 \me^{-4 A})\, ,\\
        \Phi^{f_4}&=f_4-\frac{F_0}{6 l^2}\left( y^3 j+\frac{F_0^2 y^5+l^2 y}{2}\dd y \wedge D\psi\right)+\frac{F_0y^2}{72 l^2}(F_0^2 y^4+ 3 l^2 )\, ,\\
        \Phi^{f_6}&=f_6-\frac{1}{12l^3}\left( y^3\frac{F_0^2 \me^{4A}y^2+3 l^2}{F_0^2 \me^{4A}y^2+ l^2}j\wedge j+(F_0^2  y^5+3 l^2 y)\dd y\wedge D\psi \wedge j\right)\\
        &+\frac{1}{144 l^3}\Big( 2 y^2(F_0^2 y^4+9 l^2) j +(F_0^2 y^4+9 l^2)(F_0^2 y^4+l^2) \dd y \wedge D\psi\Big)\\
        &-\frac{y}{1296 l^3}(F_0^2 y^4+9 l^2 )^2\,.
    \end{split}
\end{equation}
These polyforms will allow us to perform flux quantization and to compute the free energy.

There are additional polyforms that one can construct. For example we have that there is the closed global two-form:
\begin{equation}
    Y= j+\frac{F_0^2 y^2+l^2 y^{-2}}{2}\dd y\wedge D\psi\, ,
\end{equation}
with equivariantly closed the polyform
\begin{equation}\label{eq:PhiY}
    \Phi^{Y}=Y-\frac{1}{6y}\left(F_0^2 y^4- 3l^2\right)\, .
\end{equation}
Note that $\Phi_2^{f_4}\sim y^3 Y$. 
There is an analogous closed global four-form that we can construct, given by
\begin{equation}
  Z=j\wedge j+(F_0^2 y^2+l^2 y^{-2})\dd y\wedge D\psi\wedge j\, ,
\end{equation}
with associated polyform
\begin{equation}
    \Phi^{Z}=Z-\frac{1}{6}\left(F_0^2 y^3 -\frac{3 l^2 }{y}\right)\Big(2 j+(F_0^2 y^2+l^2 y^{-2})\dd y\wedge D\psi\Big)+\frac{(F_0^2 y^4-3 l^2)^2}{36 y^2}\, .
\end{equation}
These will turn out to be useful when the geometry contains homologically trivial two-cycles, see section \ref{sec:SE}.

\subsubsection{The \texorpdfstring{$l=0$}{l=0} case}\label{sec:l=0}
Not all the polyforms in section \ref{sec:genericpoly} are well-defined for $l=0$, in particular the Page flux polyforms. 
When $l=0$ it is easy to see that the $B$-field vanishes (i.e.\ becomes pure gauge)  and therefore the Page fluxes and Maxwell fluxes become equivalent. Moreover, it follows that both the magnetic $F_2$ and $F_6$ fluxes become trivial, leaving only $F_4$ and the Romans mass as the non-trivial fluxes. The non-trivial four-form flux (which is closed and therefore defines a conserved charge) is
\begin{equation}\label{eq:F4l0}
    \begin{split}
         F_4&=-\frac{F_0}{2}\Bigg[\frac{1}{F_0^4 \me^{4A}y^2}j\wedge j 
        -\Big(\dd(\me^{-4A}y D\psi)- y^{-1}\dd\rho\Big)\wedge \left(\frac{1}{F_0^2 }j +\frac{y^2}{2}\dd y\wedge D\psi\right)\Bigg]\,.
    \end{split}
\end{equation}
 The polyform for this flux is 
\begin{equation}\label{eq:PhiF4l0}
\Phi
^{F_4}=   F_4-\frac{F_0}{2}\Big[\frac{y}{F_0^2\me^{4A}}j+\frac{y}{2}\dd y\wedge D\psi\Big]+\frac{F_0y^2}{8}\, .
\end{equation}
The polyforms for the free energy, $Y$, and $Z$, are obtained by taking the smooth $l=0$ limit of the expressions above.

\subsection{Localization}\label{sec:loc}

Physical observables such as the free energy, flux quantization and conformal dimensions of dual operators are computed by integrating the closed polyforms from the previous subsection. Using the fixed point formulae \eqref{BVAB246}, these integrals are particularly easy to perform. From our earlier discussion \ref{sec:degeneration} there are two choices of boundary conditions we will consider. Either the R-symmetry circle shrinks at $\me^{4A}= y^2\neq0$ or the boundary condition for an O8-plane ($l=0$ only) where $\me^{2A}=y=0$. In each case, $y$ is fixed, and we use this in the localization formulae. Notice that in the $l=0$ case all the lower dimensional parts of the polyforms vanish upon substituting $y=0$. Given the form of the contributions from the boundary in \eqref{eq:BVABboundary} it is clear that for the O8-plane there are no contributions from the boundary! We emphasize that this is not generic, other boundaries would contribute, the O8-plane is special. 

Therefore when using the localization formulae we need only sum over the contributions from the shrinking R-symmetry circle. In the $l=0$ case we present, for exposition of the formulae, the general BV--AB results. 
The on-shell action reads
\begin{align}\label{Flocl0}
\mathcal{F}=&\frac{1}{24\pi^2 \ls^8} \Bigg\{-\sum_{\Sigma_0} \frac{1}{d_0}\frac{F_0^2}{20}\frac{ y^5}{\epsilon_1\epsilon_2\epsilon_3}\bigg|_{\Sigma_0} +\sum_{\Sigma_2} \frac{1}{d_2}\frac{1}{\epsilon_1\epsilon_2}\int_{\Sigma_2}\left[\frac{y^2}{4\pi}j+\frac{F_0^2 y^5}{20}\Big(\frac{c_1(L_1)}{\epsilon_1}+\frac{c_1(L_2)}{\epsilon_2}\Big)\right] \nn \\  -  &\sum_{\Sigma_4}  \frac{1}{d_4}\int_{\Sigma_4}\bigg[\frac{1}{F_0^2}
\frac{1}{(2\pi)^2\epsilon_1}\frac{1}{y}j\wedge j+\frac{y^2}{4\pi\epsilon_1^2}j\wedge c_1(L)+\frac{F_0^2 y^5}{20\epsilon_1^3}c_1(L)\wedge c_1(L)\bigg]\Bigg\}\,,
\end{align}
For $l=0$ there are only the 4-form fluxes,
which read
\begin{align}
\hspace{-.225cm}
    N^I=&\frac{1}{(2\pi\ell_s)^3}\int_{\Gamma_4^I} F_4 
    = \frac{F_0}{4\pi\ell_s^3} \left[\sum_{\Sigma_0} \frac{1}{d_0}\frac{y^2}{4\epsilon_1\epsilon_2}\bigg|_{\Sigma_0}- \sum_{\Sigma_2} \frac{1}{d_2}\int_{\Sigma_2}\left[\frac{1}{2\pi \epsilon_1}
    \frac{1}{F_0^2y}j+\frac{y^2}{4\epsilon_1^2}{c_1(L)}\right]\right]\, ,
\end{align}
where recall that $\Sigma_{0,2}$ denote fixed submanifolds inside $\Gamma_4^I$. The four-cycles $\Gamma_4^I$ are four-cycles in $M_6$, which are not entirely fixed by the action of $\xi$. On the other hand when they are entirely fixed one has
\begin{equation}
    N=\frac{1}{(2\pi\ell_s)^3}\int_{\Sigma_4} F_4 = -\frac{1}{(2\pi\ell_s)^3}
    \int_{\Sigma_4}
    \frac{1}{2F_0^3y^4}\,j\wedge j \,.
\end{equation}

There is another observable that we will compute, namely the conformal dimensions of certain BPS operators in the dual conformal field theory. These are given by the action of D2-branes wrapped on calibrated two-cycles $\Sigma_2$ in $M_6$ \cite{Bah:2018lyv} 
\begin{equation}
    \Delta(\Sigma_2)=\frac{1}{(2\pi)^2\ell_s^3}\int_{\Sigma_2} F_0\me^{2A}y\,\vol_{\Sigma_2}=\frac{1}{(2\pi)^2\ell_s^3}\int_{\Sigma_2}Y'\,,
\end{equation}
where
\begin{equation}
    Y'= F_0\Big[\frac{y}{F_0^2\me^{4A}}j+\frac{y}{2}\dd y\wedge D\psi\Big]= -2\Phi_2^{F_4}\,,
\end{equation}
with $\Phi_2^{F_4}$ the two-form part of the polyform for $F_4$ in equation \eqref{eq:PhiF4l0}.
The BV--AB formula then gives
\begin{equation}\label{ConfDim}
  \Delta(\Sigma_2)=-\frac{F_0}{2\pi\ell_s^3}\sum_{\Sigma_0}\frac{1}{d_0}\frac{y^2}{4\epsilon_1}  \,,\qquad \Delta(\Sigma_2)=\frac{1}{(2\pi)^2\ell_s^3}\int_{\Sigma_2}\frac{1}{F_0 y}j
    \, ,
\end{equation}
where the first expression holds when $\Sigma_2$ is not entirely fixed by $\xi$, and the second when it is. 
The localization of the $l\neq0$ polyforms from \ref{sec:genericpoly} follows in a similar way. Since the expressions are not particularly insightful, we do not write them out here but rather use them directly in the examples.


\section{Examples}\label{sec:examples}

In the following we specify the general formulae form the previous section on explicit examples for the topology of $M_6$. We will consider the following examples: 
\begin{enumerate}
\item[\ref{sec:RiemannnSurface}] 5d SCFTs on a Riemann surface,
\item[\ref{sec:HS2overspindle}] 5d SCFTs on a spindle,
\item[\ref{sec:HS2overB4}] 3d SCFTs from massive deformed 3d quivers,
\item[\ref{sec:S2overB4}] 3d Chern--Simons theories from M-theory on a Sasaki--Einstein manifold,
\item[\ref{sec:SE}] 3d Chern--Simons theories from suspension of a Sasaki--Einstein manifold.
\end{enumerate}

\subsection{D4-branes wrapped on 
a Riemann surface}\label{sec:RiemannnSurface}

As a first application of our proposal we study the near-horizon of the 5d SCFTs arising from $N$ D4-branes probing a type IIA background with an O8-plane and $N_f$ D8-branes compactified on a Riemann surface. The AdS$_4$ supergravity solutions were studied in \cite{Bah:2018lyv}. Recall that the 5d SCFTs are dual to the Brandhuber--Oz AdS$_6$ solutions \cite{Brandhuber:1999np} which have internal space a topological four-dimensional hemi-sphere $HS^4$.

Consider $M_6$ to be a four-dimensional hemi-sphere bundle over a Riemann surface with projection map $\pi:M_6\rightarrow \Sigma_g$. We view the hemi-sphere bundle as being embedded in the $\mathbb{R}^5_{+}$ bundle $\mathcal{L}_1\oplus \mathcal{L}_2\oplus \mathbb{R}_{+}$, with the $\mathcal{L}_i$ two complex line bundles. We can take coordinates $\{ z_1,z_2,x\}$  on $\mathbb{R}^5_+$ with $x\geq 0$ and the hemi-sphere is embedded as $|z_1|^2+|z_2|^2+x^2=1$. The north pole is then located at the point $\{ z_1=z_2=0,x=1\}$ and the boundary, which is a three-sphere, is located at $x=0$.

We take the total space to be a Calabi--Yau threefold with bundle:\footnote{Observe that in comparing with \cite{Bah:2018lyv} we take $p_{1}=-p_{\text{there}}$ and $p_{2}=-q_{\text{there}}$.}
\begin{equation}
    \mathcal{O}(-p_1)\oplus \mathcal{O}(-p_2)\rightarrow \Sigma_g\, ,
\end{equation}
where we identify $\mathcal{L}_i=\mathcal{O}({-p_i})$.
In order for this to be Calabi--Yau the degrees of the line bundles must satisfy $p_1+p_2=\chi(\Sigma_g)=2(1-g)$. We may then write the R-symmetry vector as 
\begin{equation}
    \xi=\sum_{i=1}^{2} b_i \partial_{\varphi_i}\, ,
\end{equation}
with each $\partial_{\varphi_i}$ rotating the line bundle $\mathcal{L}_i$ defined above. The $b_i$ are then directly the weights $\epsilon_i$ in the localization formulae (by definition). 
The first Chern classes of the line bundles $\mathcal{L}_i$ in terms of $p_i$ are 
\begin{equation}\label{chernpi}
    \int_{\Sigma_g}c_1(\mathcal{L}_i)=-p_i\,.
\end{equation}

We now need to consider the fixed point set of $\xi$. This is simply a copy of $\Sigma_g$ at the pole of $HS^4$, and we denote this as $\Sigma^p_g$. There is also a boundary (from the perspective of the 6d internal metric) at $y=0$ which is the boundary of the hemi-sphere and where all contributions to the localization formulae vanish. 

There are three four-cycles that we must quantize the flux over. The first one is the full four-dimensional hemi-sphere $HS^4$ itself. Performing the integration we have
\begin{equation}\label{NRiemann}
    N=\frac{1}{(2\pi\ell_s)^3}\int_{HS^4}F_4=\frac{1}{2\pi\ell_s^3}\frac{1}{b_1b_2}\frac{F_0 y_p^2}{8}\,,
\end{equation}
from which we deduce\footnote{Since we are solving for a quadratic there is of course the negative root too. The two choices give the same final results, up to a choice of orientation. It is not hard to see that they are identical after the redefinition $y\rightarrow -y$ which just interchanges the choice of pole of the hemi-sphere we are working with. }
\begin{equation}\label{fn:signy}
    y_p=4\sqrt{\frac{\pi l_s^3}{F_0}b_1b_2N}\,. 
\end{equation}
The other four-cycles, which we denote by $C^i_4$, are two-dimensional hemi-sphere bundles over the Riemann surface. The two-dimensional hemispheres are given by the embedding $HS^{2,i}\subset\mathcal{L}_i\oplus\mathbb{R}_+$ and the four-cycles are constructed by fibering these over the Riemann surface. One finds
\begin{equation}
    N_i=\frac{1}{(2\pi\ell_s)^3}\int_{C^i_4}F_4=\frac{1}{(2\pi\ell_s)^3}\left(\int_{\Sigma_g^p}-\frac{2\pi}{b_i}\frac{1}{2F_0 y_p}j+\left(\frac{2\pi}{b_i}\right)^2\frac{F_0 y_p^2}{8}p_i\right)\,,
\end{equation}
where the second term has been integrated using \eqref{chernpi}.
Moreover we find that $N_1=-p_2 N$ and $N_2=-p_1 N$. This is a result of cohomological considerations, see appendix \ref{app:cohomology}. 
Then we may use this to find an expression for the integral of $j$ over the fixed Riemann surface as
\begin{equation}
    \int_{\Sigma_g^p}j=32\sqrt{\pi^5 \ell_s^9 F_0 b_1 b_2 N^3} (b_1 p_2 + b_2 p_1)\,.  
\end{equation}
Finally the free energy is given by
\begin{align}
    \mathcal{F}&=-\frac{16 \pi^3}{(2 \pi \ell_s)^8} \frac{(2 \pi)^2}{b_1 b_2} \left(\frac{y_p^2}{24}\int_{\Sigma^p_g}j - \frac{2 \pi}{
     b_1 b_2} \frac{F_0^2 y_p^5}{240} \left(\frac{p_1}{b_1} + \frac{p_2}{b_2}\right)\right) \\ \nn
    &=- \frac{16\sqrt{2}\pi}{5n_0^{1/2}}\sqrt{b_1 b_2}(b_1 p_2 + b_2 p_1) N^{5/2}\,.
\end{align} 
Given that the Killing spinors have   weight $\tfrac{1}{2}$ under $\xi$ (in our conventions) the coefficients $b_i$ satisfy the constraint $b_1+b_2=1$, see \cite{BenettiGenolini:2023ndb} for the analogous statement for $S^4$ bundles. 
We may therefore write them in terms of a single parameter $\varepsilon$ 
\begin{equation}
    b_1=\frac{1}{2}(1+\varepsilon)\,, \quad b_2=\frac{1}{2}(1-\varepsilon)\,.
\end{equation}
Similarly, since $p_1+p_2=2(1-g)$ we may parameterize them in terms of a new variable $z$ (valid for $g\neq1$)
\begin{equation}
    p_1=(1-g)(1+\kappa^{-1} z)\, ,\quad  p_2=(1-g)(1-\kappa^{-1} z)\, ,
\end{equation}
with $\kappa$ the extrinsic curvature $\kappa=1$ for $g=1$, $\kappa=0$ for $g=1$ and $\kappa=-1$ for $g>1$. In terms of these variables the free energy reads: 
\begin{align}
    \mathcal{F}&=- \frac{8\sqrt{2}\pi}{5\kappa n_0^{1/2}}\sqrt{(1-\varepsilon^2)}(g-1)(\varepsilon z-\kappa) N^{5/2}\, .
    \end{align}
This is an \emph{off-shell} result and should be extremized for the parameter $\varepsilon$, finding
\begin{equation}
    \varepsilon_{\pm}=\frac{\kappa\pm\sqrt{\kappa^2+8 z^2}}{4z}\, .
\end{equation}
We need to keep the $\varepsilon_+$ solution only since the $\varepsilon_-$ gives a negative free energy.\footnote{One can raise the point of whether there is a metric for both of these choices of $\epsilon$. The logical claim would be that there is an obstruction to finding a metric with the $\epsilon_-$ solution. A putative way of discerning this is by studying the conformal dimensions of BPS particles obtained by wrapping D2 branes on two-cycles. From the results below (see \eqref{eq:DeltaRiemann}) we find that the $\epsilon_-$ solution gives two positive conformal dimensions and one negative. The latter is then signalling an obstruction to finding explicit metrics.}
After extremization one finds the beautiful on-shell free energy
\begin{equation}\label{FonshellRiemann}
    \mathcal{F}=\frac{2\pi (1-g) N^{5/2}}{5 \kappa n_0^{1/2}}(\sqrt{\kappa^2+8 z^2}-3\kappa)\sqrt{1-\frac{\kappa(\sqrt{\kappa^2+8 z^2}+\kappa)}{4z^2}} \,.
\end{equation}
This matches with the field theory result obtained in
\cite{Bah:2018lyv,Faedo:2021nub} after a little rewriting and gives a derivation of the off-shell gravitational block formula of \cite{Faedo:2021nub} for the Riemann surface directly in massive type IIA. 

We can also use our localization formulae to compute the conformal dimensions of certain BPS operators which correspond to D2-branes wrapping two-cycles. 
For the three obvious two-cycles in the geometry we find:
\begin{align}\label{eq:DeltaRiemann}
    &\Delta(HS^{2,1})=-\frac{F_0}{2\pi\ell_s^3}\frac{y_p^2}{4b_1}=2b_2N=(1-\varepsilon)N\,,  \nn \\ 
    &\Delta(HS^{2,2})=-\frac{F_0}{2\pi\ell_s^3}\frac{y_p^2}{4b_2}=2b_1N=(1+\varepsilon)N\,, \\ \nn
    &\Delta(\Sigma_g^p)=-\frac{1}{(2\pi)^2\ell_s^3}\int_{\Sigma_g^p}\frac{1}{F_0 y}j=-2(b_1p_2+b_2p_1)N=\frac{2(1-g)}{\kappa}(z\varepsilon-\kappa)N\,.
\end{align}
These are the general off-shell results plugging in the on-shell value of $\varepsilon=\varepsilon_+$ gives
\begin{equation}
    \begin{split}
        \Delta(HS^{2,i}) &=\left(1\mp\frac{\sqrt{\kappa^2+8 z^2}+\kappa}{4z}\right)N\,, \\ 
    \Delta(\Sigma_g^p)&=\frac{(1-g)}{2 \kappa}\left(\sqrt{\kappa^2+8 z^2}-3\kappa\right)N\,,
    \end{split}
\end{equation}
with the $\mp$ signs corresponding to $i=1,2$ respectively. The on-shell expressions match \cite{Bah:2018lyv}. Note that requiring these to be positive fixes $|z|>1$ for $\kappa=1$ and $|z|>0$ for $\kappa=-1$. The torus case can be obtained by setting $\kappa=0$ and $g=1$ in both expressions with $\tfrac{1-g}{\kappa}\rightarrow 1$. This is indeed the correct field theory constraints and it is satisfying that it pops out from gravity. 

Before we wrap up this section it is useful to write the analogous results of the conformal dimensions if we took the $\varepsilon=\varepsilon_-$ solution to the extremal problem. We would have found:
\begin{equation}
    \begin{split}
        \Delta(HS^{2,i})&=\left(1\mp \frac{\kappa-\sqrt{\kappa^2+8 z^2}}{4z}\right)N\, ,\\
        \Delta(\Sigma_g^p)&=\left(\frac{(g-1)(3\kappa+\sqrt{\kappa^2+8z^2})}{2\kappa}\right)N\, .
    \end{split}
\end{equation}
One can see that the sign of $\Delta(\Sigma_g^p)$ and $\Delta(HS^2,i)$ are different despite all being required to be of the same (positive) sign. This then gives an obstruction to the existence of metrics with $\varepsilon=\varepsilon_-$. Recall that a similar obstruction has been observed in the geometric dual of $\mathcal{I}$/c-extremization \cite{Couzens:2017nnr,Couzens:2018wnk}, it would be interesting to further refine these statements on obstructions to the existence of metrics.

\subsection{D4-branes wrapped on a spindle}\label{sec:HS2overspindle}

Having considered D4-branes wrapping a constant curvature Riemann surface we now turn our attention to D4-branes wrapping a spindle.\footnote{One can also replace the spindle with a topological disc \cite{Suh:2021aik} which are different global completions of the same local solution. We will not concern ourselves with discs in this work though the extension is straight-forward once one accepts introducing an additional boundary contribution.} A spindle is the two-dimensional weighted projective space $\Sigma=\mathbb{WCP}^1_{[n_+,n_-]}$. It is topologically a two-sphere but with conical deficit angles $2\pi(1-n_\pm)$ at the poles and first appeared in supergravity theories in \cite{Ferrero:2020laf,Ferrero:2020twa} with many other solutions in various theories appearing since. The explicit supergravity solution corresponding to D4-branes wrapped on a spindle was constructed in \cite{Faedo:2021nub}. As shown in \cite{Ferrero:2021etw} there are different types of spindle solutions distinguished by the mechanism to preserve supersymmetry: known as the \emph{twist} and \emph{anti-twist}. In the twist case setting the two orbifold parameters to be trivial, $n_+=n_-=1$, we recover a two-sphere and indeed this reduces to the results of the previous section. In \cite{Faedo:2021nub} a gravitational block formula for the free energy of the dual SCFT was conjectured which we will derive from our results. Recently this was recovered in \cite{Suh:2024asy} by using equivariant localization in 6d U$(1)^2$ gauged supergravity with an AdS$_4\times \Sigma$ ansatz. 

Similarly to the previous section, we consider a four-dimensional hemi-sphere orbibundle over a two-dimensional space, in this instance a spindle. As before we view the hemi-sphere via the embedding $HS^4\subset \mathcal{L}_1\oplus\mathcal{L}_2\oplus \mathbb R_+\subset \mathbb R^5$. In the twist class the fibration of the two line orbibundles over the spindle
$\mathcal{O}(-p_1)\oplus \mathcal{O}(-p_2)\rightarrow \Sigma$ 
is Calabi-Yau, which enforces the condition $p_1+p_2=n_++n_-$. To impose the anti-twist one should instead take $p_1+p_2=n_+-n_-$. This may be unified by introducing the sign $\sigma=\pm1$ such that $p_1+p_2=n_++\sigma n_-$ with the twist given by $\sigma=1$ and the anti-twist given by $\sigma=-1$.

The main difference in the localization in comparison with the Riemann surface analysis resides in the fact that the R-symmetry vector can now also rotate the spindle, and therefore the R-symmetry vector takes the form 
\begin{equation}
    \xi=\sum_{i=1}^{2} b_i \del_{\varphi_i}+\varepsilon\del_{\varphi_0}\, ,
\end{equation}
where $\del_{\varphi_i}$ rotate the $\mathcal{L}_i$ and $\del_{\varphi_0}$ the spindle. Therefore the weights at the fixed points are identified with $\epsilon_i=b_i$, and $\epsilon_3=\mp\varepsilon/n_\pm$. The fixed point set is qualitatively very different from before as we now have isolated fixed points (located at the poles of the spindle and hemi-sphere) rather than a fixed two-dimensional surface. Consequently, even though the setups seem closely related, the localization analysis is fairly different as we shall see now.

Let us first quantize the flux. We get a copy of $HS^4/\mathbb Z_{n_\pm}$ at each pole of the spindle through which we can compute the fluxes\footnote{This is the same flux equation as \eqref{NRiemann} but doubled and with an orbifold factor $d=n_\pm$. Note that the subscript $p$ there denoted the pole of the hemi-sphere, while now the $\pm$ denote the poles of the spindle. These fixed points are still at the pole of the hemisphere. To be fully precise we should use the subscript $p,\pm$, but we drop the $p$ for ease of notation.}
\begin{equation}
    N_\pm=\frac{1}{(2\pi\ell_s)^3}\int_{HS^4/\mathbb Z_{n_\pm}}F_4=\frac{1}{2\pi\ell_s^3}\frac{1}{n_\pm}\frac{1}{b_1^\pm b_2^\pm}\frac{F_0 y_\pm^2}{8}\,.
\end{equation}
Moreover the homologies of these cycles are related, see appendix \ref{app:cohomology} and \cite{Boido:2022mbe}. We have
\begin{equation}
    n_+N_+=n_-N_-\equiv N\,,
\end{equation}
such that\footnote{We see that there is a sign ambiguity with the roots of $y_{\pm}$. Without loss of generality we can take $y_{+}$ to be positive and then we introduce a relative sign between the two roots. With hindsight comparing with the literature \cite{Faedo:2021nub} we identify this sign with $\sigma$, however we emphasize that this is not dictated by the above.\label{foot:peppapig}} 
\begin{equation}
    y_+=4\sqrt{\frac{\pi l_s^3}{F_0}b_1^+ b_2^+ N}\,, \quad  y_-=4\sigma \sqrt{\frac{\pi l_s^3}{F_0}b_1^- b_2^- N}\,.
\end{equation} 
The flux quantization through other cycles is not needed to obtain the free energy (this is to be contrasted with the Riemann surface computation), and so we postpone performing this for the moment. 

We can directly plug the expressions for $y_\pm$ back into the free energy. This equation is quite different from the section before since we are now localizing on isolated fixed points rather than fixed surfaces,   
\begin{align}
\mathcal{F}&=-\frac{1}{24\pi^2 \ls^8}\frac{F_0^2}{20} \left(\frac{1}{n_+}\frac{ y_+^5}{b^+_1 b^+_2 (-\varepsilon/n_+)}+\frac{1}{n_-}\frac{ y_-^5}{b^-_1 b^-_2 (\varepsilon/n_-)} \right)\\ \nn
&=\frac{32\sqrt{2}\pi}{15n_0^{1/2}}\frac{(b_1^+b_2^+)^{3/2}-\sigma(b_1^-b_2^-)^{3/2}}{\varepsilon}N^{5/2}\,.
\end{align}
This derives the conjectured gravitational block formula in \cite{Faedo:2021nub} directly in massive type IIA. The ambiguity in which gluing to pick is then related to the sign of the fixed points $y_{\pm}$, this is similar to what happens in the explicit solutions in the 5d and 4d cases \cite{Ferrero:2021etw, Couzens:2021cpk}. It would be interesting to understand any obstructions to finding explicit metrics for these different choices.

To proceed with the extremization it is useful to  identify\footnote{Notice that we have $2b_i^{\pm}=\Delta_{i}^{\pm}$ when comparing with the notation of \cite{Faedo:2021nub}.}
\begin{equation}\label{eq:spindlebs}
    \begin{split}
        b_i^{\pm}=b_i +\frac{\varepsilon}{2} \left(\pm\frac{p_i}{n_+n_-}+\frac{n_+-\sigma n_-}{2n_+n_-}\right)\, ,\quad b_1+b_2=1\,,
    \end{split}
\end{equation}
which can be read off from the results in \cite{Boido:2022mbe} or \cite{Faedo:2021nub}. More concretely one can realize this by noting that this solves the conditions: 
\begin{equation}
    b_1^{\pm}+b_{2}^{\pm}=1\pm \frac{\varepsilon}{n_{\pm}}\, ,\qquad b_i^{+}-b_{i}^{-}=\frac{p_i}{n_+n_-}\varepsilon\, .
\end{equation}
The former conditions set the overall charge of the holomorphic volume form to be 1 as required and the latter are equivalent to (3.24) of \cite{Boido:2022mbe} which define the gluing of the different patches used to build $M_6$. 

One should now extremize this off-shell free energy over $\varepsilon$ and $b_i$ subject to the constraint \eqref{eq:spindlebs}. Since this has been done already in \cite{Faedo:2021nub} (see also \cite{Suh:2024asy}) we will not give the extremization explicitly and refer the interested reader to those works. 
Upon extremization this gives the free energy for the compactifcation of the 5d $\mathcal{N}=1$ USp$(2N)$ gauge theory with $N_f=8-n_0$ massless hypermultiplets in the fundamental representation and one hypermultiplet in the antisymmetric representation of USp$(2N)$ on a spindle.

Having recovered the off-shell gravitational block formula for the free energy we return to the quantization of the fluxes. There are other four cycles through which we can quantize the flux, and while they were not needed for the free energy computation, they are interesting quantities in their own right. The cycles we want to consider are the total space of the $HS^{2,i}\subset\mathcal{L}_i\oplus\mathbb{R}_+$ bundle over the spindle and denoted by $C_4^i$. The localization formula gives
\begin{equation}
    N_i=\frac{1}{(2\pi\ell_s)^3}\int_{C^i_4}F_4=\frac{1}{2\pi\ell_s^3}\frac{F_0}{8}\left(\frac{1}{n_+}\frac{y^2_+}{b_i^+(-\varepsilon/n_+)}+\frac{1}{n_-}\frac{y^2_-}{b_i^-(\varepsilon/n_-)}\right)\,,
\end{equation}
and after simplifying we find
\begin{equation}
    N_1=\frac{b_2^--b_2^+}{\varepsilon}N=\frac{p_2}{n_+n_-}N\,,\quad
    N_2=\frac{b_1^--b_1^+}{\varepsilon}N=\frac{p_1}{n_+n_-}N\,.
\end{equation}
This is the same result as in the M5-brane case \cite{BenettiGenolini:2023ndb} and can be explained using cohomological arguments similar to those in appendix \ref{app:cohomology}. 

Before finishing the section we can study certain BPS operators in the dual field theory which correspond to D2-branes wrapping calibrated cycles. There are two types of calibrated cycle, one is the two cycle consisting of the spindle at the pole of the hemi-sphere and the second type are the two-dimensional hemispheres $HS^{2,i}$ discussed above at a pole of the spindle. This is similar to the M5-brane case as explained in \cite{Bomans:2024mrf}. For the D2-branes wrapping a copy of the spindle at the pole of the $HS^4$, the off-shell conformal dimension is
\begin{equation}
    \Delta(\Sigma^p)=-\frac{F_0}{2\pi\ell_s^3}\left(\frac{1}{n_+}\frac{y_+^2}{4(-\varepsilon/n_+)}+\frac{1}{n_-}\frac{y_-^2}{4(\varepsilon/n_-)} \right)=-2\frac{b_1^-b_2^--b_1^+b_2^+}{\varepsilon}N\,.
\end{equation}
For the second type of BPS operator we have
\begin{equation}
    \Delta(HS^{2,i}_{\pm})=\frac{F_0}{2\pi \ls^3}\frac{y_{\pm}^2}{4 b_{i}^{\pm}}=2\frac{b_{1}^{\pm}b_{2}^{\pm}}{b_{i}^{\pm}}N\, .
\end{equation}
An obvious obstruction to finding explicit metrics is to require these conformal dimensions to be positive, however it is clear that since the $y_{\pm}$ always appear squared this will not fix the sign ambiguity noted in footnote \ref{foot:peppapig}. We leave understanding these points to future work.

\subsection{\texorpdfstring{$HS^2$}{HS2} bundle over \texorpdfstring{$B_4$}{B4}}\label{sec:HS2overB4}

We now turn our attention to hemi-sphere bundles over a four-dimensional base with the requirement that $M_6$ is complex, see section 4 of \cite{Passias:2018zlm}. We will impose that there is an O8-plane which implies that we consider $M_6$ as the total space of an $HS^2$ bundle over a 4-dimensional base $B_4$, with $HS^2$ the two-dimensional hemi-sphere\footnote{We refer interchangeably to the hemi-sphere as a disc (compact with boundary) since they are topologically identical. The hemi-sphere notation makes explicit the choice of boundary condition.}
\begin{equation}
    HS^2\hookrightarrow M_6\rightarrow B_4\, .
\end{equation}
Recall from section \ref{sec:degeneration} that this implies that we necessarily need to take $l=0$.
While $B_4$ could in principle be any compact manifold, for $M_6$ to be a complex manifold $B_4$ is either K\"ahler-Einstein or the product of two Riemann surfaces which can be seen by using the arguments of \cite{Gauntlett:2004zh,Apostolov2001}.
Explicit solutions of this type have been constructed in \cite{Passias:2018zlm} and  further studied in \cite{Colombo:2023fhu}. This section recovers and generalizes their results. 

We assume that the vector field $\xi$ rotates just the $HS^2$ fibre and leaves $B_4$ fixed. The known solutions are constructed by fibring the hemi-sphere bundle using the anti-canonical line bundle $\mathcal{L}$ over $B_4$. Note that since we are considering the hemi-sphere in our localization computations, we have a spacetime with a boundary. As explained earlier, the general formulae as written in section \ref{sec:loc} do not contain such boundary contributions, however since the boundary is for an O8 plane located at $y=0$ and the U$(1)$ action acts freely on the boundary, the boundary terms do not contribute and we may use those results. Recall that this requires the warp factor $\me^{2A}$ to degenerate in a certain way, and is what fixes the boundary conditions.  Our localization locus consists of a single fixed point at the pole of the hemi-sphere, denoted $y_p$ and the vanishing boundary contribution at $y=0$. The fixed point at the pole is actually a fixed copy of $B_4$ with normal bundle $\mathcal{L}$.

Following \cite{BenettiGenolini:2023ndb} we define 
\begin{equation}\label{eq:candnB4}
    c_\alpha=\int_{\Gamma_\alpha} j\,, \quad n_\alpha=\int_{\Gamma_\alpha}  c_1(\mathcal{L})\,,
\end{equation}
where $\Gamma_\alpha$ form a basis for the 2-cycles of the copy of $B_4$ at the pole, and $\mathcal{L}$ is its the normal line bundle.
Moreover the weight at the pole is $\epsilon=1$ the sign fixed by our choice of $c_1(\mathcal{L})$.
Using this we can quantize the flux. The first possibility is to consider four-cycles $C_4^\alpha$ which are the total space of the $HS^2$ bundle over $\Gamma_\alpha$, giving $N_\alpha$ 
\begin{equation}
    N_\alpha=\frac{1}{(2\pi\ell_s)^3}\int_{C_4^\alpha} F_4=\frac{F_0}{4\pi\ell_s^3}\left[\frac{1}{2\pi}\frac{1}{F_0^2y_p}c_\alpha-\frac{y_p^2}{4}n_\alpha \right]\,,
\end{equation}
from which we can deduce an expression for $c_\alpha$ in terms of $N_\alpha$ and $n_\alpha$
\begin{equation}\label{calpha}
    c_\alpha=\frac{\pi F_0y_p}{2}(F_0y_p^2n_\alpha-16\pi\ell_s^3N_\alpha)\,.
\end{equation}
Then we can also quantize the flux through the full fixed $B_4$ giving $N_p$ 
\begin{equation}
\begin{split}
    N_p&=\frac{1}{(2\pi\ell_s)^3}\int_{B^4_p} F_4=-\frac{1}{(2\pi\ell_s)^3}\frac{1}{2F_0^3 y_p^4}\langle c,c \rangle \\&= \frac{1}{2}\la N,n\ra-\frac{F_0 y_p^2}{64\pi\ell_s^3}\la n,n\ra-\frac{4\pi\ell_s^3}{F_0 y_p^2}\la N,N\ra
\,,
\end{split}
\end{equation}
where we used \eqref{calpha} in the second line. To keep the equations succinct it is useful to define the bracket $\langle c,c\rangle\equiv I_{\alpha\beta}c_\alpha c_\beta$ and similarly for other quantities in the following. Here $(I_{\alpha\beta})$ is the inverse of the intersection form for $B_4$, see appendix \ref{app:cohomology}.
Finally as shown in appendix \ref{app:cohomology}, the cycles are not independent such that we get some topological relation between the various fluxes which reads\footnote{When comparing with \cite{Passias:2018zlm,Colombo:2023fhu} the parameter $M$ is related to having a non-zero value of $\sigma$ in the notation of \cite{Passias:2018zlm}.}
\begin{equation}
    N_p=-\langle N,n\rangle+M\,.
\end{equation}
Combining the two previous relations for $N_p$, we can solve for $y_p$, giving  the solutions
\begin{equation}
    y_p=4\sqrt{\frac{\pi\ell_s^3}{F_0}\frac{3\la N,n\ra-2M\pm \sqrt{(3\la N,n\ra-2M)^2-\la N,N\ra\la n,n\ra}}{\la n,n\ra}}\, .
\end{equation}
Inserting these results for the $y_p$ into the free-energy localized form we obtain two positive solutions
\begin{align}
\label{FB4}
    \mathcal{F}&=-\frac{1}{24\pi^2 \ls^8}\left[\frac{1}{(2\pi F_0)^2 y_p}\la c,c\ra-\frac{y_p^2}{4\pi}\la c,n\ra+\frac{F_0^2 y_p^5}{20}\la n,n\ra\right]\nonumber  \\ 
    &=
    \frac{2y_p}{3\ell_s^2}\la N,N\ra-\frac{F_0^2y_p^5}{1920\pi^2\ell_s^8}\la n,n\ra \\ 
    &=    \frac{16\sqrt{2}\pi}{15n_0^{1/2}}\frac{1}{\la n,n\ra^{3/2}}
    \sqrt{3\la N,n\ra-2M\pm \sqrt{(3\la N,n\ra-2M)^2-\la N,N\ra\la n,n\ra}}\nonumber \\ &\times\left(3\la N,N\ra \la n,n\ra +(3\la N,n\ra-2M) \left(3\la N,n\ra\pm \sqrt{(3\la N,n\ra-2M)^2-\la N,N\ra\la n,n\ra}\right)\right)\,,\nn
\end{align}
where the first line is just the localization \eqref{Flocl0}, then we replaced $c_\alpha$ in the second line, and $y_p$ in the last line, giving our final result.
In summary this is the result for a general topology on $B_4$. Picking a topology amounts to specifying the $N,n$-brackets and we will consider different examples shortly in the following and compare with known solutions. Note that the parameter $M$ generalizes the results in \cite{Colombo:2023fhu}. We also note that there are two different choices, both giving a well defined free energy for suitable constraints on the bracket. The comments about obstructions may, once again be repeated virtually verbatim.  

Finally we can also compute the conformal dimensions using \eqref{ConfDim}. We can either consider a D2 brane wrapping the fibre $HS^2$ itself giving
\begin{equation}
    \Delta(HS^2)=\frac{F_0}{2\pi\ell_s^3}\frac{y_p^2}{4}=\frac{2}{\la n,n\ra}\left(3\la N,n\ra-2M\pm \sqrt{(3\la N,n\ra-2M)^2-\la N,N\ra\la n,n\ra}\right)\,,
\end{equation}
or alternatively wrapping $\Gamma_\alpha$ which results in
\begin{align}
    \Delta(\Gamma_\alpha)&=\frac{1}{(2\pi)^2\ell_s^3}\frac{c_\alpha}{F_0 y_p}\\ \nn &=\frac{2}{\la n,n\ra}\left( (3\la N,n\ra-2M) n_\alpha + (\la n,n\ra\pm\sqrt{(3\la N,n\ra-2M)^2-\la N,N\ra\la n,n\ra}) N_\alpha\right)\, .
\end{align}

\vspace{3mm}

\subsubsection{K\"ahler--Einstein base}
For our first example we consider $B_4=KE_4^+$ to be a positively curved K\"ahler--Einstein four-manifold. Since we have taken a K\"ahler--Einstein base the fluxes are proportional to the Chern numbers $N_\alpha=k n_\alpha$. Then we have
\begin{equation}
    \la n,n\ra= \bar M\,, \quad \la N,n\ra= k \bar M\,, \quad \la N,N\ra= k^2 \bar M\,,
\end{equation}
where $\bar M$ is a topological invariant (the integral of the Ricci form squared), and we further define
\begin{equation}
    k=\frac{N}{h}\,.
\end{equation}
Plugging these in the general expression for the free energy and setting the flux number $M=0$ straightforwardly gives
\begin{equation}
    \mathcal{F}=
    \frac{32\sqrt{2}\pi}{5n_0^{1/2}}
    (3\pm2\sqrt{2})\bar M\left(\frac{N}{h}\right)^{5/2}\,.
\end{equation}
This matches exactly the results obtained in \cite{Colombo:2023fhu}.
Additionally the conformal dimensions read
\begin{equation}
    \Delta(HS^2)=\frac{2(3\pm2\sqrt{2})}{h}N\,,\qquad
    \Delta(\Gamma_\alpha)=\frac{4(1\pm\sqrt{2})}{h}Nn_\alpha\,,
\end{equation}
which have not appeared in the literature previously. Note that for the $-$ sign that $\Delta(\Gamma_{\alpha})<0$, one suspects that this then indicates an obstruction to finding a metric in this case, whilst the other case seems pathology free.

\subsubsection{Product base}
The second option is to take $B_4$ to be the product of two Riemann surfaces: $B_4=\Sigma_{g_1}\times\Sigma_{g_2}$. Then the brackets are
\begin{equation}
    \la n,n\ra= 2\chi_1\chi_2\,, \quad \la N,n\ra= \chi_1N_2+\chi_2N_1\,, \quad \la N,N\ra= 2N_1N_2\,.
\end{equation}
It is then straightforward to insert these in the general formula \eqref{FB4} to obtain the free energy. We secretly assumed that our base had a positive overall curvature which implies that at least one of the Riemann surfaces should be a sphere for $B_4$ to be positively curved and our computations to apply. 

In particular considering the product of two spheres, $\chi_1=\chi_2=2$ and setting $M=0$, we can parametrize the fluxes in terms of a single variable $N_1=(1+z)N$, $N_2=(1-z)N$, with $|z|<1$ and obtain 
\begin{equation}
    \mathcal{F}=
    \frac{32\pi}{5n_0^{1/2}}
    \sqrt{3\pm\sqrt{8+z^2}}(\sqrt{8+z^2}\pm(2+z^2))N^{5/2}\,.
\end{equation}
Again this result matches with the one of \cite{Colombo:2023fhu}. We also have
\begin{equation}\begin{split}
    &\Delta(HS^2)=\left(3\pm\sqrt{8+z^2}\right)N\,,\\
    &\Delta(S^2_1)=2\left(2-z\pm\sqrt{8+z^2}\right)N\,, \quad \Delta(S^2_2)=2\left(2+z\pm\sqrt{8+z^2}\right)N\,,
\end{split}
\end{equation}
where the copies of the $S^2$ from the base are taken at the pole of the $HS^2$ fibre. 
Similarly to the K\"ahler--Einstein case, with the minus sign, the $\Delta(S^2_i)$ become negative and one would suspect that this indicates an obstruction to finding explicit metrics for this solution.


\subsection{\texorpdfstring{$S^2$}{HS2} bundle over \texorpdfstring{$B_4$}{B4}}\label{sec:S2overB4}

Next let us consider an $S^2$ bundle over a four-dimensional base $B_4$. In this case we will no longer force the presence of an O8-plane which accounts for considering an $S^2$ rather than $HS^2$ and allows us to consider the $l\neq0$ class. Moreover we are no longer able to gauge away $l$, as one can do when both $F_0$ and $l$ are non-zero, and we will see that we in fact need to extremize over it. 

To make things more tractable we will restrict to the massless case and make some comments about the massive case at the end of this section. Recall that solutions of this form can be constructed by reducing certain Sasaki--Einstein manifolds from M-theory. Additionally one can consider the massive deformation of these theories by turning on a non-trivial Romans mass, see \cite{Lust:2009mb,Petrini:2009ur,Aharony:2010af,Tomasiello:2010zz}.

Since we are considering an $S^2$ bundle over a four-dimensional base the fixed point locus corresponds to the base $B_{4}$ at the two poles of the sphere. We use the anti-canonical line bundle $\mathcal{L}$ of $B_4$ to fibre the $S^2$ over $B_4$. We assume that the R-symmetry vector $\xi$ acts only on the $S^2$ such that the fixed point set is precisely copies of $B_4$ at the north and south pole of the sphere, which we call $B_4^{N,S}$. The normal bundles are then
\begin{equation}
    L^{N}=\mathcal{L}^{-1}\, ,\qquad L^{S}=\mathcal{L}\,.
\end{equation}

 As explained in the previous section, albeit for the hemi-sphere, there are two types of two-cycles to consider. One is the $S^2$ fibre and the other is a two-cycle in $B_4$ at either of the poles. We denote by $b_2$ the 2nd Betti number on $B_4$ and define $\Gamma_{\alpha}^{N,S}$ to be the two-cycles on $B_4$ at the north and south pole of the sphere respectively. These cycles are not independent, following \cite{BenettiGenolini:2023ndb} (see also appendix \ref{app:cohomology}), we have
\begin{equation}\label{eq:homS2}
    [\Gamma_{\alpha}^{S}]-[\Gamma_{\alpha}^{N}]=n_{\alpha}[S^2_{\text{fibre}}]\in H^2(M_6,\mathbb{Z})
\end{equation}
where
\begin{equation}
  n_{\alpha}\equiv \int_{\Gamma_{\alpha}}c_1(\mathcal{L})\in \mathbb{Z}\, .  
\end{equation}
The four-cycles have a similar structure, they are the copies of $B_4$ at the two poles, $B_4^{N,S}$ and four-cycles constructed by taking
the sphere bundle over one of the cycles $\Gamma_{\alpha}$ which we denote by $C_{\alpha}$. Again these cycles are not independent and we have the relation
\begin{equation}\label{eq:S24dhom}
    [B_4^{S}]-[B_4^{N}]=\sum_{\alpha=1}^{b_2}n_{\alpha}[C_{\alpha}]\,.
\end{equation}

Analogously to the definition of $c_{\alpha}$ in \eqref{eq:candnB4}, we define
\begin{equation}
    c_{\alpha}^{N,S}\equiv \int_{\Gamma_{\alpha}^{N,S}}j\, .
\end{equation} Using the homological relation \eqref{eq:homS2} with the equivariantly closed form $\Phi^Y$, \eqref{eq:PhiY} we find the constraint
\begin{equation}\label{eq:cfixS2B4}
  c_{\alpha}^{S}-c_{\alpha}^{N} =\frac{l^2 \pi}{\epsilon}\left(\frac{1}{y^N} -\frac{1}{y^{S}}\right)n_{\alpha}\, ,
\end{equation}
where $\pm\epsilon$ is the weight of $\xi$ at the north and south pole respectively, and the charge of the Killing spinor fixes $\epsilon=1$. This is analogous to the constraint \eqref{eq:candnB4} in the previous section, though different due to the bundle structure. One could also integrate the equivariantly closed four-form $\Phi^Z$ over the four-cycles and use the homology relation \eqref{eq:S24dhom} and would reassuringly find that this is satisfied given \eqref{eq:cfixS2B4}.

Next let us consider the quantization of the fluxes. In the massless case it is not difficult to see that the four-form flux vanishes and we can choose the gauge in which the $B$ field vanishes too. We therefore have two types of fluxes to quantize, the two-form $f_2$ and the six form $f_6$.\footnote{There is a subtlety here that we are sweeping under the carpet slightly. Since we may fix a gauge where $B=0$ one can use the Maxwell fluxes to quantize the fluxes. Rather than computing a whole new set of polyforms we will set $F_0=0$ in our Page flux polyforms and since the polyforms use the gauge in which $B=0$ if we set $F_0=0$ these will be equivalent. This still leaves possible the option of turning on a closed term in $B$ and computing the Page flux with this choice. Notice that this choice is made in \cite{Benini:2011cma} for example. This ambiguity leaves a gauge freedom which we will encounter later and is related to the parameter $l$ not being fixed by anything.}
Let us first quantize the two-form flux. To avoid a proliferation of flux integers called $n$ we denote these quantized fluxes associated to the two-form $f_2$, by $k$. In the field theory there is a relation between these and the Chern--Simons levels of the dual field theory, see for example \cite{Benini:2011cma, Tomasiello:2010zz}. We have
\begin{equation}
    \begin{split}
    k_2~&=\frac{1}{2\pi\ls}\int_{S^2_{\text{fibre}}} f_2=\frac{l}{2\epsilon \ls}\left[\frac{1}{y_N}-\frac{1}{y_S}\right]\, ,\\
        k^{N,S}_{\alpha}&=\frac{1}{2\pi\ls}\int_{\Gamma_{\alpha}^{N,S}} f_2=\frac{c_{\alpha}^{N,S}}{2 l \ls \pi}\, .
    \end{split}
\end{equation}
We see that we may rewrite the constraint \eqref{eq:cfixS2B4} as
\begin{equation}\label{eq:cins2B4new}
    c_{\alpha}^S-c_{\alpha}^{N}= 2 l \ls \pi k_2 n_{\alpha} \quad\Leftrightarrow \quad k_{\alpha}^S-k_{\alpha}^N= k_2 n_{\alpha}\, .
\end{equation}
The quantization conditions fixes the $c_{\alpha}^{N,S}$ in terms of the flux parameters $k_{\alpha}^{N,S}$ (associated to D6-branes wrapping four-cycles or KK-monopoles in the M-theory setup). Note that using the constraint \eqref{eq:cins2B4new} one sees that for integer $n_{\alpha}$, $k_2$, and $k_{\alpha}^{N}$ necessarily $k_{\alpha}^{S}\in \mathbb{Z}$ as required. We use the definition of $k_2$ to fix one of the roots, without loss of generality we solve for $y_N$:
\begin{equation}
    y_N=\frac{l y_S}{l+2 \ls k_2 y_S\epsilon}\, .
\end{equation}
Next let us compute the quantization of $f_6$. We have
\begin{equation}
    \begin{split}
    N&=-\frac{1}{(2\pi\ls)^5}\int_{M_6}f_6\\
    &=\frac{l^4 k_2^2\pi^2(y_S-y_N)+l^2 k_2\epsilon \pi(c_N y_N^2-c_S y_S^2)+\epsilon^2(c_S^2 y_S^3-c_N^2 y_N^3)}{64 l^3 \epsilon^3 \pi^4 \ls^5}\, .
    \end{split}
\end{equation}
We need to solve this for $y_S$ however we should be smart about this. We redefine the parameter $l$ as
\begin{equation}
    l= -L y_S k_2 \ls \,,
\end{equation}
with $L$ the new free parameter. We find
\begin{equation}
    N=\frac{y_S^2(\langle k^N,k^N\rangle L^2+\langle k^S,k^S\rangle(L-2)^2+\langle k^N,k^S\rangle L(L-2))}{8 L (L-2)^3\ls^4 k_2 \pi^2}\, ,
\end{equation}
where for ease of presentation we have written $N$ in terms of $k^S_\alpha$ rather than $n_\alpha$ using \eqref{eq:cins2B4new} and set $\epsilon=1$ as required by supersymmetry.
We may finally compute the free energy, to save presenting the messy intermediate results we just give the result
\begin{equation}
    \mathcal{F}=\frac{4\pi }{3}\sqrt{\frac{2 k_2^{3} L^{3}(L-2)^{3}}{\langle k^N,k^N\rangle L^2+\langle k^S,k^S\rangle (L-2)^2+\langle k^N,k^S\rangle L (L-2)}}N^{3/2}\, .
\end{equation}
We now need to extremize over the remaining free parameter $L$. It is not hard to see that the condition to solve is the cubic\footnote{Another extremal value is $L=0$ or $L=2$ however these will lead to a vanishing free energy and so we can discount them. }
\begin{equation}\label{eq:LS2}
 \langle k^N,k^N\rangle L^2(2L-1) +\langle k^S,k^S\rangle (L-L)^2(2L-3)+2\langle k^N,k^S\rangle L(2-3 L+L^2)=0\, .
\end{equation}
One can solve this simply using Mathematica, obtaining a result in terms of cubic roots of the parameters $k_2$, $k^{N}_{\alpha}$, and $n_{\alpha}$. It is not very instructive to present the roots, though one can show that a single root exists since the discriminant of \eqref{eq:LS2} is negative.

Having computed the general on-shell free energy we can compare to the reduction of $Y^{p,k}(\text{KE}_4)$ to massless type IIA. We will follow the conventions of \cite{Martelli:2008rt} for $Y^{p,k}(\text{KE}_4)$. The free energy is 
\begin{equation}
\mathcal{F}=\sqrt{\frac{2 \pi^6}{27\Vol(Y^{p,k}(\text{KE}_4))}}N^{3/2}\, .
\end{equation}
We can extract out the volume from \cite{Martelli:2008rt},
\begin{equation}
    \Vol\big(Y^{p,k}(\text{KE}_4)\big)=\Vol(\text{KE}_4)\frac{2\pi^2}{3
    \cdot 4^4} \frac{x_2-x_1}{p(x_2-1)(1-x_1)}(x_2^3-x_1^3)\, ,
\end{equation}
where the $x$'s are solutions of
\begin{equation}
    \begin{split}
        0&=3 p^3 x_1^3+2 p^2(6 b-5p)x_1^2+p(18 b^2-28 b p+11 p^2)x_1+4(3 b^3+4 b p^2-6 b^2 p-p^2)\, ,\\
        0&=3 p^3 x_2^3+2 p^2(p-6 b) x_2^2+p(18 b^2-8 b p +p^2)x_2+4 b(3 b p-3 b^2-p^2)\, .
    \end{split}
\end{equation}
The dictionary between our results and those in \cite{Martelli:2008rt} are\footnote{Note that the north pole here is the south pole in \cite{Martelli:2008rt} and vice versa. },
\begin{equation}
k_2\equiv p,  \quad k_{\alpha}^{S}=\frac{k}{h} n_{\alpha}\equiv b n_{\alpha}\, , \quad k_{\alpha}^{N}=(b-p) n_{\alpha}\, ,  \quad \Vol(\text{KE}_4)=\frac{\pi^2}{2}\langle n,n\rangle\, ,
\end{equation}
where $h=\text{gcd}(n_{\alpha})$. We find a perfect match. 

Observe that we have found an extremal problem for a class of Sasaki--Einstein manifolds using equivariant localization. This is of the same spirit as the use of equivariant localization for odd dimensional manifolds in \cite{Goertsches:2015vga}. One uses a second U$(1)$ to reduce along, then localizes with the remaining U$(1)$. It would be interesting to consider more general reductions, for example one could write a seven-dimensional Sasaki--Einstein manifold as a fibration of $L^{a,b,c}$ over a round two-sphere and then reduce along one of the non R-symmetry directions. One would then not obtain a K\"ahler--Einstein base as we have assumed above. 
One can also consider turning on a non-trivial Romans mass. Our localized integrals can be used to study this case however one finds the need to solve difficult algebraic equations which we were unable to solve without resorting to numerics. This is certainly an interesting class of solutions and it would be interesting to see if there is a change of parameters which makes the problem tractable.

\subsection{Suspension of a Sasaki--Einstein manifold}\label{sec:SE}

An interesting class of 3d $\mathcal{N}=2$ SCFTs can be constructed by considering D2-branes probing the suspension of a Sasaki--Einstein manifold in the presence of D8-branes. The dual field theories are 3d Chern--Simons theories with the level specified by the Romans mass, see for example \cite{Fluder:2015eoa,Tomasiello:2010zz}. We will first consider the case where the Sasaki--Einstein manifold is the round five-sphere $S^5$ before considering a more general Sasaki--Einstein manifold. Note that this class keeps both $l$ and $F_0$ non-zero.

\subsubsection{Round five-sphere}

Consider first the suspension of the round five-sphere, $\mathcal{S}(S^5)$. There are two fixed points at the poles of the suspension, which we call the North and South pole respectively. We may write the R-symmetry vector as
\begin{equation}
    \xi=\sum_{i=1}^{3} \epsilon_i \partial_{\varphi_i}\, ,
\end{equation}
where the $\partial_{\varphi_i}$ generate the U$(1)^3$ toric action of the $S^5$. Consider the three linearly embedded four-spheres $S_i^4\subset \mathcal{S}$ which are invariant under the action of $\xi$. These are all trivial in homology such that we necessarily have
\begin{equation}
    0=\int_{S_i^4}\Phi^Z=\frac{(2\pi)^2\epsilon_i}{\epsilon_1\epsilon_2\epsilon_3}\left[\frac{(F_0^2 y_N^4-3l^2)^2}{36 y_N^2}+\frac{(F_0^2 y_S^4-3l^2)^2}{36 y_S^2}\right]\, .
\end{equation} 
Since the two terms are positive definite each must vanish individually and therefore
\begin{equation}\label{eq:ysinS(S5)}
    |y_N|=|y_{S}|=\left(\frac{3 l^2}{F_0^2}\right)^{1/4}\, .
\end{equation}
Note that we can consider three linearly embedded two spheres also: $S_i^2\subset \mathcal{S}(S^5)$ which are invariant under $\xi$ and are again trivial in homology. Integrating the closed two-form $Y$ over these cycles gives
\begin{equation}
    0=\int_{S_i^2}\Phi^Y=-\frac{\pi}{3\epsilon_i}\left[\frac{1}{y_N}(F_0^2 y_N^4-3l^2)-\frac{1}{y_S}(F_0^2 y_N^S-3l^2)\right]\, ,
\end{equation}
which is  true after application of \eqref{eq:ysinS(S5)}. 

Next consider the quantization of the fluxes. We only have a six-cycle on which we may integrate our fluxes and therefore we introduce the quantum number $N$ counting the number of D2-branes as
\begin{equation}
    N=\frac{1}{(2\pi \ls)^5}\int_{M_6}f_6=\frac{(2\pi)^3}{1296l^3(2\pi\ls)^5\epsilon_1\epsilon_2\epsilon_3}\Big[y_N(F_0^2 y_N^4+9l^2)^2-y_S(F_0^2 y_S^4+9l^2)^2\Big]\, .
\end{equation}
Restricting to a positive $N$ we fix $y_N=-y_S>0$. The quantization of $N$ and the Romans mass imposes
\begin{equation}
\ls^{-1}= \frac{2^{1/9}3^{7/18}n_0^{1/9}N^{2/9}\pi ^{1/3}(\epsilon_1\epsilon_2\epsilon_3)^{2/9}}{l^{1/3}}\, ,\qquad F_0=\frac{n_0}{2\pi \ls}\, .
\end{equation}
It is now trivial to use our results to compute the free energy, finding
\begin{equation}
    \begin{split}
        \mathcal{F}&=\frac{(2\pi)^3}{240\epsilon_{1}\epsilon_{2}\epsilon_{3}}\frac{16\pi^3}{(2\pi\ls)^8}\Big[y_N(F_0^2 y_N^4+5l^2)-y_S(F_0^2y_S+5l^2)\Big]\\
        &=\frac{9}{5}2^{1/3}3^{1/6}n_0^{1/3}N^{5/3}\pi (\epsilon_{1}\epsilon_{2}\epsilon_{3})^{2/3}\, .
    \end{split}
\end{equation}
We can now extremize the above functional over the $\epsilon_i$ subject to the constraint $\sum_{i=1}^{3}\epsilon_i=1$, one finds the symmetric solution
\begin{equation}
    \epsilon_i=\frac{1}{3}\, . 
\end{equation}
Inserting this into the off-shell free energy we obtain:
\begin{equation}
\mathcal{F}=\frac{2^{1/3}3^{1/6}n_0^{1/3}\pi^3}{5 \Vol(S^5)^{2/3}}N^{5/3}\, ,
\end{equation}
where $\Vol(S^5)=\pi^3$. This is indeed the correct result for the free energy of D2-branes probing the suspension of the round five-sphere. 

Notice that the extremization that we are performing is nothing other than extremizing the volume of the five-sphere. In order to rewrite this in a more amenable form we note that the normalization of the Reeb vector field of the $S^5$ in comparison to our localizing Killing vector is such that the spinor has charge $3$ under the Reeb and 1 under $\xi$. Therefore in order to compare with the canonically normalized Reeb vector we should identify
\begin{equation}
    b_{i}=3\epsilon_{i}\, ,
\end{equation}
where $b_{i}$ are the mixing parameters of the Reeb vector field. With these conventions the off-shell volume of the $S^5$ is
\begin{equation}
    \Vol(S^5)[b_i]=\frac{\pi^3}{b_1 b_2 b_3}\, ,
\end{equation}
subject to the constraint $\sum_{i=1}^{3}b_i=3$.
It therefore follows, in the $S^5$ case, that we have recovered that a-maximization = F-maximization \cite{Fluder:2015eoa} since the off-shell free energy can be written as
\begin{equation}
    \mathcal{F}=\frac{2^{1/3}3^{1/6} n_0^{1/3}\pi^3}{5 (\Vol(S^5)[b_i])^{2/3}}N^{5/3}\, ,
\end{equation}
and extremizing over $b$ is equivalent to maximizing the `a'-central charge of the 4d parent theory.

\subsubsection{Quasi-regular Sasaki--Einstein}

Having recovered the $\mathcal{S}(S^5)$ case we turn our attention to studying the more general $\mathcal{S}(\text{SE}_5)$ case. We restrict to the quasi-regular Sasaki--Einstein case in the following. The Reeb vector field is once again aligned with the U$(1)_R$ vector field $\xi$ which we use to localize, with the 
factor of proportionality once again $3$. Recall a basic fact of Sasaki--Einstein geometry that the Reeb vector field has constant square norm. This implies that the action generated by the Reeb vector field is locally free on the suspension away from the poles. Hence, the fixed point contribution is entirely from the singular points at the two poles of the suspension. Since, for a generic base, the poles of the suspension are badly singular one cannot use the previous localization formulae with the line bundle decomposition and one needs to use the more general result \eqref{eq:BVABgen}. 

Consider now two-cycles on the base, in $\mathcal{S}(\text{SE}_5)$ these are trivial in homology and therefore we find, analogously to the $S^5$ case that 
\begin{equation}\label{eq:ysinS(SE5)}
    |y_N|=|y_{S}|=\left(\frac{3l^2}{F_0^2}\right)^{1/4}\, .
\end{equation}
As in the previous case there are no two-cycles nor four-cycles over which we need to integrate the fluxes and we are left with a single flux integral to perform. We have
\begin{equation}
\begin{split}
    N&=\frac{1}{(2\pi\ls)^2}\int_{\mathcal{S}(\text{SE}_5)}f_6\\
    &=\frac{1}{1296(2\pi\ls)^5}\left[ \frac{y_N(F_0^2 y_N^4+9l^2)^2}{e(\mathcal{N}_N)}+
 \frac{y_S(F_0^2 y_S^4+9l^2)^2}{e(\mathcal{N}_S)}\right]\\
    &=\frac{1}{9(2\pi\ls)^5}\left[\frac{y_N}{e(\mathcal{N}_N)}+ \frac{y_S}{e(\mathcal{N}_S)}\right]\, .
    \end{split}
\end{equation}
We have kept the normal bundles at the two poles arbitrary for the moment. The quantization conditions become
\begin{equation}
     \ls^{-1}=\frac{2\cdot 3^{7/18}\pi n_0^{1/9}N^{2/9}}{\mathcal{E}}\,,\quad F_0=\frac{n_0}{2\pi\ls}\, ,
\end{equation}
where
\begin{equation}
    \mathcal{E}=\frac{1}{e(\mathcal{N}_N)}-\frac{1}{e(\mathcal{N}_{S})}\, .
\end{equation}
We may similarly compute the free energy which gives
\begin{equation}
    \begin{split}
        \mathcal{F}&=\frac{\pi^3}{15(2\pi \ls)^8}\left[\frac{y_N (F_0^2 y_N^4+5l^2)}{e(\mathcal{N}_N)}+\frac{y_S(F_0^2 y_S^4+5l^2)}{e(\mathcal{N}_S)}\right]\\
        &=\frac{8\pi^3 y_N}{15(2\pi\ls)^8} \left[\frac{1}{e(\mathcal{N}_N)}-\frac{1}{e(\mathcal{N}_S)}\right]\,,
    \end{split}
\end{equation}
and inputting the quantization conditions we have
\begin{equation}
\mathcal{F}=\frac{72\times 3^{1/6} n_0^{1/3}\pi^3}{5\mathcal{E}^{2/3}}  N^{5/3} \, .
\end{equation}
We now want to understand what $\mathcal{E}$ is computing. Let us take $\mathcal{N}_N=-\mathcal{N}_S$ for simplicity. Then for a Calabi--Yau cone we have that\footnote{One can make this more precise by using the results of \cite{Martelli:2006yb,Goertsches:2015vga,Boyer:2017jwa}. One can understand the factors of $3$ due to the same normalization of the Reeb vector field in the previous $S^5$ example, 3 of the 2's due to the formula for the volume and one factor of the 2's since $\mathcal{E}$ double counts the volume when $\mathcal{N}_N=\mathcal{N}_S$. }
\begin{equation}
 \mathcal{E}= 2^4 3^3 \Vol(\text{SE}_5)\, ,
\end{equation}
and we obtain the final result
\begin{equation}
    \mathcal{F}=\frac{2^{1/3}3^{1/6}\pi^3}{5\Vol(\text{SE}_5)^{2/3}}n_0^{1/3}N^{5/3}\, .
\end{equation}
Strictly this is an off-shell result and one needs to vary the volume $\Vol(\text{SE}_5)$ with respect to the Reeb vector field. We have therefore recovered in the general quasi--Regular case that a-maximization $=$ F-maximization \cite{Fluder:2015eoa}.

We note that there seems to be an interesting option to have a suspension where the poles of the cone have different singularities. One can view this as having a flop between the two poles. It is not clear that such a metric exists but it would be interesting to study whether this is obstructed.


\section{Conclusions and future directions}\label{sec:conclusion}

Using equivariant localization we have constructed the geometric dual of F-maximization in massive type IIA. There is a rich structure of topologies for the internal manifolds $M_6$, beyond those for the closely related AdS$_5$ solutions in M-theory \cite{Gauntlett:2004zh} recently studied using localization in \cite{BenettiGenolini:2023kxp,BenettiGenolini:2023ndb}, that we have discussed. One of the key differences of our work with theirs is the presence of boundary terms in the internal space. For a subset of examples we have introduced boundary conditions associated to the presence of an O8-plane. Rather than the localizing circle shrinking at this locus it remains of finite size (in the internal manifold) and therefore leads to a boundary contribution. For the O8-plane we saw that there is no contribution to the free energy, flux quantization and conformal dimensions of a class of BPS operators, however this is not generic and other choices of defect branes, which are used to cap off the space, would give contributions.  

Motivated by this observation, by using equivariant localization, we may be able to study more complicated brane intersections which are used to find a compact internal space. The general AdS$_4$ solutions studied here can also be capped off with an O4-plane or O6-plane, \cite{Passias:2018zlm}, it would be interesting to extend our analysis to these examples too.  Moreover, this could open up avenues for studying more general irregular punctures, those of type 3, in the holographic duals of Argyres--Dougles theories recently studied in \cite{Bah:2021mzw,Couzens:2022yjl,Couzens:2023kyf,
Bah:2022yjf,Bomans:2023ouw}. Finding explicit supergravity solutions seems exceedingly difficult, yet if one can understand concretely the degeneration structure at such a singularity one can utilize a similar logic to this paper. 

As we have emphasized repeatedly throughout, though the equivariant localization may allow for a particular choice of topology it is not immediate that explicit metrics can be found realizing these topologies. It is therefore interesting to consider obstructions to such metrics in an analogous way to how the Futaki invariant gives an obstruction to the existence of Sasaki--Einstein metrics, \cite{Gauntlett:2006vf}. It is natural to conjecture that requiring that the conformal dimensions of wrapped D2-branes are all positive gives rise to such an obstruction, however it is unclear whether this is sufficient or necessary. One could therefore consider studying more refined observables for example the eigenvalue density of the matrix model associated to the dual field theory, see for example \cite{Boido:2023ojv,Farquet:2013cwa}, and flavour central charges associated to the defect branes \cite{Freedman:1998tz,Chang:2017mxc,Bah:2018lyv}.

Observe that as a by-product of our construction we have constructed the consistent truncation of the general AdS$_4$ solutions to four-dimensional Einstein--Maxwell supergravity.\footnote{One should check that this preserves supersymmetry but this should be automatic by construction.}  To see this clearly one should gauge the R-symmetry vector $D\psi$ with the 4d graviphoton, and in the polyforms for the fluxes wedge with the graviphoton field strength so that the form degree of the total polyform is simply the form degree of the top component.
One could then imagine using the polyforms constructed here and the polyforms for 4d Einstein--Maxwell \cite{BenettiGenolini:2023kxp} to construct polyforms for the full 10d solution. In this way one could study black holes in these massive type IIA backgrounds directly and make contact with the field theory results in \cite{Hosseini:2017fjo}.  

Note that though our supergravity solutions preserve supersymmetry there is no requirement to actually have any supersymmetry to use equivariant localization, rather one just needs some symmetry. There are often non-supersymmetric sister AdS solutions of certain classes of supersymmetric AdS solutions. For AdS$_7$ in massive type IIA there are the non-supersymmetric sister solutions of \cite{Apruzzi:2019ecr} for example, and for the AdS$_4$ Sasaki--Einstein solutions in M-theory there are various different non-susy families, for example the Englert solutions \cite{Englert:1982vs} and the Pope--Warner solutions \cite{Pope:1984bd,Pope:1984jj}. There are in fact non-supersymmetric solutions which are sister solutions of the supersymmetric solutions studied here \cite{Koerber:2010rn}. Given that the solutions have a very similar structure, in particular a U$(1)$ symmetry, one could apply equivariant localization techniques to these non-supersymmetric cases too. 

So far all the supergravity equivariantly localized solutions have used an even-dimensional localizing space. Notice that in section \ref{sec:S2overB4} we have actually localized over an odd-dimensional space. The construction uses an additional circle to first reduce on and produce an even-dimensional space over which one can now localize. There are various tools, beyond this reduction, that one may use to localize over odd-dimensional spaces \cite{Boyer:2017jwa,Goertsches:2015vga}. It would be interesting to extend these techniques to these odd-dimensional examples. Moreover one could further extend to including higher derivative corrections and therefore go beyond the large $N$ limit from supergravity. We hope to return to some of these future directions soon.


\section*{Acknowledgments}

The authors are grateful to Pieter Bomans, Achilleas Passias,  James Sparks and Alessandro Tomasiello for insightful discussions. We are also thankful to both James Sparks and Achilleas Passias for comments on the draft. CC thanks the Oxford Mathematical Institute for support. AL is supported by the Palmer scholarship in Mathematical Physics of Merton college. 
\\

CC dedicates this work to his grandfather, Alan Ernest Couzens, who sadly passed away on the 29th February 2024 at the age of 99. He was a wonderful Granddad, with whom I have many fond memories. Granddad was always very supportive and interested in my career in academia and would often ask me many interesting questions about the latest snippet of physics or mathematics that he had read up on and my own papers. He was one of the most influential people in getting me interested in mathematics and physics and I owe him a lot for all the love and support he gave me my entire life. He is deeply missed and I hope I can continue to make him proud.

\appendix



\section{Some cohomology and homology}\label{app:cohomology}

In the main text we have considered a number of different fiber bundles with 2d hemi-sphere and 4d hemi-sphere fibers. In this section we will describe some of the various homology relations that we have used. Many of the results can be extracted from \cite{Bott:1982xhp} and slight modifications, but for clarity we discuss these results and their applications to our setup. For a clear exposition of $S^2$ and $S^4$ bundles see appendix C of \cite{BenettiGenolini:2023ndb}.

Consider a six-manifold $M_6$ which is the total space of a disc bundle over a four-manifold $B_4$ with projective map $\pi:M_6\rightarrow B_4$. We consider the disc bundle inside the $\mathbb{R}^3_+$ bundle $\mathcal{L}\oplus\mathbb{R}_+ \rightarrow B_4$ where $\mathcal{L}$ is a complex line bundle. We may introduce coordinates $(z,x)$ on $\mathbb{R}^3_+$, with $x>0$ and the disc defined by $|z|^2+x^2=1$.

We may define the Thom class for the bundle. Let $\rho(r)$ be a radial function which is $1$ at the centre of the disc and 0 on the boundary. This then makes $\dd\rho=\rho'(r)\dd r$ a bump form on $\mathbb{R}$ with total integral 1. Further, define the global angular form on the disc to be $\eta$. Then the Thom class may be written as\footnote{ We thank \textit{Topology Boy} for clarifying this and other points in this appendix with us. }
\begin{equation}
    \Phi^{\text{Thom}}=\frac{1}{2\pi}\dd \left(\rho(r)\eta\right)\, .
\end{equation}
This is compactly supported in the vertical direction, is closed and integrates to 1 when restricted to the fiber. Let $s:B_4\rightarrow M_6$ be the zero section, then we have
\begin{equation}
    s^* \Phi^{\text{Thom}}=-e\,,
\end{equation}
with $e$ the Euler class of the bundle \cite{Bott:1982xhp}. It then follows that the Poincar\'e dual of $B_4$ is
\begin{equation}
    \Psi_p\equiv\text{PD}[B_4]=\Phi^{\text{Thom}}\, .
\end{equation}
Let $\Gamma_{\alpha}$ be a basis for the free part of $H_2(B_4,\mathbb{Z})$ and let $\{\Psi_{\alpha}\}$ be the Poincar\'e duals of this basis: $\Psi_{\alpha}=\text{PD}[\Gamma_{\alpha}]\in H^2(B_4)$. Further let the dimension of $H_2(B_4,\mathbb{R})$ be $b_2\equiv\text{dim}(H_2(B_4,\mathbb{R}))$. Given any closed two-form, $\tau$ on $B_4$ we may expand it in terms of this basis via 
\begin{equation}
    \tau= \sum_{\alpha=1}^{b_2} \hat{C}_{\alpha}\Psi_{\alpha}\,.
\end{equation}
Alternatively we may expand by defining the constants $C_{\alpha}$ where
\begin{equation}
    C_{\alpha}\equiv \int_{\Gamma_{\alpha}}\tau\, .
\end{equation}
Using the definition of the Poincar\'e dual we may rewrite this as 
\begin{equation}
    C_{\alpha}=\int_{B_4}\tau\wedge \Psi_{\alpha}=\sum_{\beta=1}^{b_2}Q_{\alpha\beta}\hat{C}\, ,
\end{equation}
where $Q$ is the intersection form
\begin{equation}
    Q_{\alpha\beta}\equiv \int_{B_4}\Psi_{\alpha}\wedge \Psi_{\beta}=[\Gamma_{\alpha}]\cap[\Gamma_{\beta}]\, .
\end{equation}
Recall that this is an integer-valued unimodular symmetric matrix. It follows that we may write
\begin{equation}
    \int_{B_4}\tau\wedge \tau =\sum_{\alpha,\beta=1}^{b_2}Q_{\alpha\beta}\hat{C}_{\alpha}\hat{C}_{\beta}=\sum_{\alpha,\beta=1}^{b_2}I_{\alpha\beta}C_{\alpha}C_{\beta}\, ,
\end{equation}
where $I=Q^{-1}$. This allows us to expand the first Chern class of a line bundle over $B_4$ as
\begin{equation}
    c_1(\mathcal{L})=\sum_{\alpha=1}^{b_2} \hat{n}_{\alpha}\Psi_{\alpha}\, ,
\end{equation}
with $\hat{n}_{\alpha}\in\mathbb{Z}$ which is used in section \ref{sec:HS2overB4}.
It is also useful to define the pairing $\langle\cdot\,,\cdot\rangle$ for two-forms. Let $\tau$ and $\omega$ be two two-forms with expansions:
\begin{equation}
    \tau= \sum_{\alpha=1}^{b_2} \hat{C}_{\alpha}\Psi_{\alpha}\, ,\quad \omega= \sum_{\alpha=1}^{b_2} \hat{N}_{\alpha}\Psi_{\alpha}\, ,
\end{equation}
then
\begin{equation}
    \int_{B_4}\tau\wedge \omega=\sum_{\alpha,\beta=1}^{b_2}Q_{\alpha\beta}\hat{C}_{\alpha}\hat{N}_{\beta}=\sum_{\alpha,\beta=1}^{b_2}I_{\alpha\beta}C_{\alpha}N_{\beta}\equiv\langle C, N\rangle\, .
\end{equation}
The pairing is used in section \ref{sec:HS2overB4} to simplify expressions.

Consider now a closed form-form on $M_6$. Using the Thom class we can expand the four-form $F_4$ as
\begin{equation}
    F_4=\Phi^{\text{Thom}}\wedge\sum_{\alpha=1}^{b_2} \hat{N}_{\alpha}\Psi_{\alpha}+ M \widetilde{\text{vol}}(B_4)\, ,
\end{equation}
where $\widetilde{\text{vol}}(B_4)$ is normalized to integrate to  1 over $B_4$. 
Note that whereas the contribution from the Thom class vanishes on the boundary of the disc, the second term does not and signifies that additional branes are present along the boundary. We therefore find that
\begin{equation}
\begin{split}
  (2\pi \ls)^3  N&=\int_{B_4} F_4\\
  &=M+\hat{N}_{\alpha}\int_{B_4}s^*(\Phi^{\text{Thom}})\wedge \Psi_{\alpha}\\
  &=M - \hat{N}_{\alpha}\int_{B_4} e\wedge \Psi_{\alpha}\\
  &=M-\langle N,n\rangle\, .
  \end{split}
\end{equation}

Next consider two-cycles in $M_6$ defined by taking $\Gamma_{\alpha}\subset B_4$ evaluated at the centre of the disc. By the Thom isomorphism these have Poincar\'e duals
\begin{equation}
\text{PD}[\Gamma_{\alpha}]=\pi^*\Psi_{\alpha}\wedge \Phi^{\text{Thom}}\in H^4(M_6)\, .
\end{equation}
Integrating this over $B_4$ at the pole we have
\begin{equation}
    \int_{B_4} \pi^*\Psi_{\alpha}\wedge \Phi^{\text{Thom}}= n_{\alpha}\, ,
\end{equation}
and therefore we find the homology relation
\begin{equation}
    [\Gamma_{\alpha}]=n_{\alpha}[HS^2_{\text{fibre}}]\, .
\end{equation}
Note that this holds on closed forms which vanish on the boundary of the disc.





\bibliographystyle{JHEP}

\bibliography{AdS4D2}

\end{document}